\documentclass[10pt, twocolumn]{IEEEtran}
\usepackage{graphicx}
\usepackage{subfigure}
\usepackage{amsmath}
\usepackage{amsthm}
\usepackage{amssymb}
\usepackage{algorithm}
\usepackage{algorithmicx}
\usepackage{algpseudocode}
\usepackage[T1]{fontenc}
\usepackage{color}
\usepackage{times}
\usepackage{cite}
\usepackage{setspace}

\begin{document}
\title{Gain without Pain: Recycling Reflected Energy from Wireless Powered RIS-aided Communications}
\author{Hao Xie, \IEEEmembership{Student Member,~IEEE,} Bowen Gu, \IEEEmembership{Student Member,~IEEE,} Dong Li,~\IEEEmembership{Senior Member,~IEEE,} Zhi Lin, Yongjun Xu, ~\IEEEmembership{Senior Member,~IEEE} 
	\thanks{H. Xie, B. Gu, and D. Li are with the School of Computer Science and Engineering, Macau University of Science and Technology, Avenida Wai Long, Taipa, Macau 999078, China (e-mails: 3220005631@student.must.edu.mo, 21098538ii30001@student.must.edu.mo, dli@must.edu.mo).}
	\thanks{Z. Lin is with the College of Electronic Engineering, National University of Defense Technology, Hefei 230037, China, is also with the School of Computer Science and Engineering, Macau University of Science and Technology, Taipa, Macau 999078, China (e-mail: linzhi945@163.com).}	
	\thanks{Y. Xu is with the School of Communication and Information Engineering, Chongqing University of Posts and Telecommunications, Chongqing 400065, China (e-mail: xuyj@cqupt.edu.cn).}
} 

\maketitle
\pagestyle{empty}
\thispagestyle{empty}

\begin{abstract}
In this paper, we investigate and analyze energy recycling for a reconfigurable intelligent surface (RIS)-aided wireless-powered communication network. As opposed to the existing works where the energy harvested by Internet of things (IoT) devices only come from the power station, IoT devices are also allowed to recycle energy from other IoT devices. In particular, we propose group switching- and user switching-based protocols with time-division multiple access to evaluate the impact of energy recycling on system performance.  Two different optimization problems are respectively formulated for maximizing the sum throughput by jointly optimizing the energy beamforming vectors, the transmit power, the transmission time, the receive beamforming vectors, the grouping factors, and the phase-shift matrices, where the constraints of the minimum throughput, the harvested energy, the maximum transmit power, the phase shift, the grouping, and the time allocation are taken into account. In light of the intractability of the above problems, we respectively develop two alternating optimization-based iterative algorithms by combining the successive convex approximation method and the penalty-based method to obtain corresponding sub-optimal solutions.  Simulation results verify that the energy recycling-based mechanism can assist in enhancing the performance of IoT devices in terms of energy harvesting and information transmission. Besides, we also verify that the group switching-based algorithm can improve more sum throughput of IoT devices, and the user switching-based algorithm can harvest more energy.

%
%

\end{abstract}


\IEEEpeerreviewmaketitle

\section{Introduction}
The development and popularization of Internet of Things (IoT) technologies has greatly facilitated the information-processing capabilities of IoT devices, a variety of IoT devices such as controllers, sensors, actuators, and communication devices substantially promotes the implementation of remote adaptation and configuration, digital twins, and device collaborative operation \cite{a1}. However, the limited energy supply has become a major bottleneck in restricting the communication performance of IoT devices due to the increased energy consumption and the limited battery size of IoT devices. Frequent battery replacement and redeployment will greatly increase the cost, and is also extremely difficult in extreme environments\cite{a2}. How to charge these energy-limited devices in a green and low-carbon way is an imminent issue.
\subsection{Related Works}
Radio-frequency (RF) energy harvesting, which can supply wireless energy for IoT devices by converting received RF signals, stands out due to its convenience, stability, and controllability compared to intermittent ambient energy harvesting technologies,  such as solar, wind, and tidal \cite{b1}. Wireless-powered communication network (WPCN) is one of the main paradigms evolved from RF energy harvesting\cite{b2}, where a dedicated power station supplies energy for IoT devices. WPCN intelligently utilizes the broadcast nature of RF radios to make one-to-many long-distance charging possible \cite{b3}. Despite these promising results, WPCNs still suffer from low efficiency due to the following facts: Firstly, lowly efficient wireless energy transfer (WET) and wireless information transfer (WIT) result in limited downlink harvested energy and lower uplink transmission rates of IoT devices. Secondly, the harvest-then-transmit (HTT) mode has a lower spectral efficiency since the WIT in WPCNs is based on the traditional time-division multiple access (TDMA) manner in most existing works. Finally, IoT devices suffer from the ``doubly near-far'' effect in WPCNs, which leads to an imbalance between the energy harvested and the energy consumed by IoT devices.

There have been many research efforts that have been paid to overcome the above challenges. For instance, one of the solutions for addressing the inefficiency of WET and WIT in WPCNs is that the hybrid access point (HAP) operates in a full-duplex (FD) mode\cite{c1}, where the HAP transmits and receives over the same time and the same frequency band, thus potentially yielding double the spectral efficiency of the half-duplex. However, the FD technology suffers from serious self-interference at the HAP, which is the critical challenge \cite{c2}. Fortunately, current self-interference cancellation techniques are able to suppress the self-interference power to the noise floor\cite{c3}. On the other hand, one way to mitigate the ``doubly-near-far'' effect is to deploy base stations on the unmanned aerial vehicle (UAV), a wireless-powered ALOHA network was proposed in \cite{c4} to maintain the line of sight transmission of energy and information. Another way to overcome the ``doubly-near-far'' effect is to investigate the max-min fairness, i.e., focusing on the IoT device with the worst performance and maximizing its performance. The max-min rate optimization problem was investigated in \cite{c5} by using zero-forcing-based space-division multiplexing under the transmit power constraint. Different from the fairness performance study in \cite{c5}, a cooperative transmission protocol was considered in \cite{c6} to achieve the fairness transmission, including the WET phase, the single WIT phase, and the cooperative WIT phase, where each user harvested energy during the WET phase, broadcasted its own information to the HAP and other users during the single WIT phase, and all users jointly transmitted information to the HAP during the cooperative WIT phase.

Recently, reconfigurable intelligent surface (RIS), also known as intelligent reflecting surface (IRS), which consists of a large number of passive and low-power reflecting units, has been proposed to improve the performance for wireless communication networks in terms of transmission reliability and efficiency \cite{d1}. Specially, the RIS is able to adjust the phase and amplitude of the incident signal since each reflecting unit can be configured independently, thus achieving active customization of wireless environment \cite{d4}. Besides, the reflected signals can be superimposed or eliminated, thus enhancing the desired signal strength or suppressing the harmful interference \cite{d5}. Several investigations on the RIS-aided communication have been explored from various network scenarios, such as device-to-device networks \cite{e1}, millimeter-wave networks \cite{e2}, UAV-aided networks \cite{e3}, non-orthogonal multiple access (NOMA)-aided networks\cite{e4}, heterogeneous networks\cite{e5}, cognitive networks\cite{e6}, satellite communication\cite{e7}, etc., and different performance metrics of RIS-aided networks have also been analyzed, such as power consumption\cite{f1}, transmission rate\cite{f2}, energy efficiency\cite{f3}, outage probability\cite{f4}, fairness transmission \cite{f5}, phase-shift feedback \cite{f6}, etc. Most of the existing works have applied the RIS to WIT, which, however, ignore the enormous benefits of the RIS to WET. Passive beamforming of the RIS can compensate for severe penetration loss, and create local hot spots or charging zones, which substantially enhances the efficiency of downlink WET and uplink WIT. So far, few works have been published on RIS-aided WPCNs \cite{g1,g2,g4,g5,g6}. For example, three beamforming configurations, which consist of fully dynamic RIS beamforming, partially dynamic RIS beamforming, and static RIS beamforming, were proposed in \cite{g1} to strike a balance between the system performance, signaling overhead, and implementation complexity. The authors in \cite{g2} introduced the RIS to enhance WET, and the sum throughput and the total power consumption were balanced by maximizing the total energy efficiency. The integration of RIS and NOMA can also boost the performance of WPCNs \cite{g4}, i.e., the HAP first transmits information to the information user via the NOMA or applies the HTT mode to charge the energy-limited user, and then users apply the harvested energy to transmit information to the HAP based on NOMA. To mitigate the impacts of the ``doubly-near-far'' effect, the authors in \cite{g5} adopted the three-phase transmission-based user cooperation strategy in \cite{c6} to achieve the max-min fairness for RIS-aided WPCNs. Different from the existing works on passive RIS-aided WPCNs,  the active RIS was introduced in \cite{g6}, where three dynamic RIS beamforming, including user-adaptive RIS beamforming, uplink-adaptive RIS beamforming, and static RIS beamforming, were proposed to alleviate user unfairness.

\subsection{Motivation and Contributions}
Although extensive studies have been conducted on WPCNs in recent years, there are still some open problems. To be specific, the harvested power at the receiver side, despite requiring dedicated power sources for supply, is prone to the varying channel conditions, thus resulting in an unstable received signal strength and prolonging the accumulated power process for the HTT. One way to combat these problems is to deploy power sources with a high power level, which is, however, costly particularly for a proliferated number of power stations. Thus, a natural question arises: how to enhance the signal strength in WPCNs without requiring an extra power budget?

In this paper, we explore the reflected energy recycling to solve the above problem for the RIS-aided WPCN, where a multiple-antenna HAP with the FD mode serves multiple single-antenna IoT devices, and the RIS is deployed in the cellular network to assist the HAP and IoT devices for downlink WET and uplink WIT. Besides, it is allowed that IoT devices can not only harvest the energy transmitted by the HAP, but also recycle the energy from the uplink WIT from other devices. It is noted that \cite{h2,h3,h4} are related to this work regarding energy recycling. However, only self-interference recycling of the FD relay was considered in \cite{h2}, and the problem of energy recycling among multiple devices is not addressed. Although multiple devices were considered in \cite{h3,h4}, the gain of the RIS on energy harvesting and information transmission has not been evaluated. To bridge the above gaps, we propose a user switching-based protocol and a group switching-based protocol, respectively, which are, to the best of the authors' knowledge, not presently available in existing works. The main contributions of this paper are summarized as follows.
\begin{itemize}
	\item For the user switching-based protocol, we consider a periodic transmission protocol to supply sufficient energy and guarantee stable transmission for IoT devices, where IoT devices harvest energy in all phases except its own WIT phase, and perform uplink WIT in a TDMA manner. For the group switching-based protocol, one time slot is divided two phases and IoT devices are divided into two groups based on whether they perform uplink WIT or downlink WET, where one group performs uplink WIT and the other group performs downlink WET during the first phase,  then the roles of two groups are switched during the second phase. For the two different protocols, we formulate different optimization problems to compare their performance. Specifically, the sum throughput of all IoT devices is maximized by jointly optimizing the energy beamforming vectors, the transmit power, the receive beamforming vectors, the grouping factors, and the phase-shift matrices subject to the minimum throughput constraint, the harvested energy constraint, the maximum transmit power constraint, the phase-shift constraint, the grouping constraint, and the time allocation constraint.
	\item  To solve these two intractable and non-convex problems, we respectively propose two efficient algorithms. In particular, two alternating optimization (AO)-based iterative algorithms by combining the successive convex approximation (SCA) method and the penalty-based method are proposed to obtain the corresponding sub-optimal solutions. Furthermore, we derive the closed-form solutions for the receive beamforming vectors by applying the minimum-mean-square-error (MMSE) and the maximal-ratio-combining (MRC), respectively. To gain more insight, we compare the performance in terms of transmit power and energy harvesting under the two protocols under average time division and equal phase shift.  
	\item Simulation results verify that the energy recycling-based mechanism enables IoT devices to harvest more energy, thereby improving the system throughput. We also find that the group switching-based protocol can improve more sum throughput of IoT devices, which is more suitable for scenarios with high throughput requirements, and the user switching-based protocol can harvest more energy, which is more suitable for scenarios with power sources with a low power level.
\end{itemize}

The rest of this paper is structured as follows. The system model is presented in Section  \uppercase\expandafter{\romannumeral2}. The group switching-based protocol and the corresponding algorithm are presented in Section \uppercase\expandafter{\romannumeral3} and \uppercase\expandafter{\romannumeral4}, respectively. Section \uppercase\expandafter{\romannumeral5} and Section \uppercase\expandafter{\romannumeral6} introduce the user switching-based protocol and the corresponding algorithm, respectively. Section \uppercase\expandafter{\romannumeral7} gives the performance comparison. Simulation results are presented in Section \uppercase\expandafter{\romannumeral8}. The paper is concluded in Section \uppercase\expandafter{\romannumeral9}.

\textit{Notations:} In this paper, $\mathbb{E}^{N \times M}$ and $\mathbb{C}^{N \times M}$ denote the statistical expectation and the set of all $N \times M$ complex-valued matrices, respectively. $\boldsymbol{{\rm I}}_M$ and $\boldsymbol{0}$ denote the $M \times M$ identity matrix and all-zero matrix, respectively. $x$, $\boldsymbol{\rm x}$, and $\boldsymbol{\rm X}$ denote the scalar, the vector, and the matrix, respectively. The distribution of a circularly symmetric complex Gaussian random variable with mean $\mu$ and variance $\sigma^2$ is denoted by $\mathcal{CN}(\mu,\sigma^2)$. ${\rm diag}(\cdot)$, ${\rm Tr}(\cdot)$, and ${\rm Rank}(\cdot)$ mean the diagonalization operation, the trace, and the rank, respectively. $\boldsymbol{\rm X}^H$ and $[\boldsymbol{{\rm X}}]_{i,j}$ stand for the conjugate transpose and the $(i,j)$-th element of matrix $\boldsymbol{\rm X}$, respectively. $\boldsymbol{\rm X}\succeq \boldsymbol{0}$ indicates that $\boldsymbol{\rm X}$ is a positive semidefinite matrix. $\|\boldsymbol{\rm X}\|_2$ and $\|\boldsymbol{\rm X}\|_*$ denote the spectral norm and the nuclear norm of matrix $\boldsymbol{\rm X}$, respectively. 


\section{System Model}
We consider an RIS-aided WPCN, where an HAP equipped with $M_{\rm T}$ transmit antennas and $M_{\rm R}$ receive antennas serves $K$ single-antenna IoT devices. The HAP operates in the FD mode and an RIS with $N$ passive reflecting units is deployed in the cell for assisting the downlink WET and uplink WIT of IoT devices. Define $\forall m,j, k\in\mathcal{K}=\{1,...,K\}$ and $\forall n\in\mathcal{N}=\{1,...,N\}$ as the indexes of IoT devices and reflecting units, respectively. We assume that IoT devices are energy-limited devices with an energy harvesting circuit so that they can harvest downlink energy transmitted by the HAP and the uplink signals. Furthermore, the quasi-static flat-fading channel is assumed, where the channel state information remains constant within a channel coherence frame but may change for different frames. In what follows, we propose two TDMA protocols to evaluate the performance of the RIS-WPCN, i.e., the group switching-based protocol and the user switching-based protocol.
\begin{figure*}[!t]
	\centering
	\includegraphics[width=5.2in]{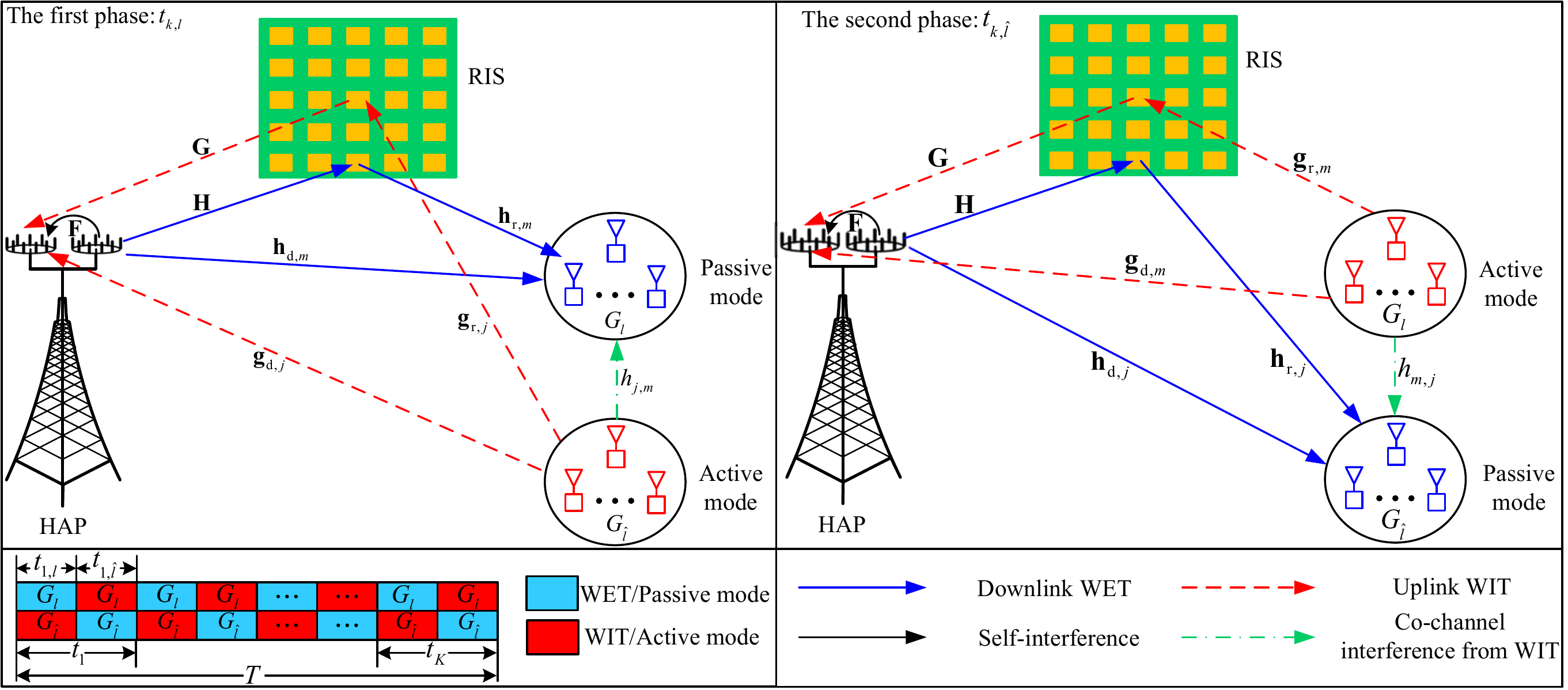}
	\caption{An RIS-aided WPCN with group switching.}
	\label{fig1}
\end{figure*}
\section{The Group Switching-based Protocol}
We first investigate and analyze the group switching-based protocol. Specifically, IoT devices are divided into two groups based on whether they perform uplink WIT or downlink WET. As shown in Fig. \ref{fig1}, assuming that there are multiple IoT devices communicating with the HAP, where one part of IoT devices is classified in the group $G_l$ to receive downlink energy signals from the HAP via a passive mode while another part of IoT devices is classified in the group $G_{\hat l}$ to transmit uplink information signals to the HAP during the first phase $t_{1,l}$ via an active mode. Then, the roles of groups $G_{ l}$ and $G_{\hat l}$ are switched during the second phase $t_{1,\hat{l}}$.

\subsection{The $l$-th phase}
During the downlink WET, the received signal of the $m$-th IoT device in phase $l$ of time slot $k$ is given by  
\begin{equation}\label{eq1}
\begin{array}{l}
y_{m,k,l}^{\rm DL}=\underbrace{(\boldsymbol{\rm h}_{{\rm d},m}^H+\boldsymbol{\rm h}_{{\rm r}, m}^H\boldsymbol{\rm \Theta}_{k,l}\boldsymbol{\rm H})\boldsymbol{\rm v}_{k,l}s_{k,l}}_{\textrm{Energy\ signal}}\\
+\underbrace{\sum\limits_{j\in G_{\hat{l}}}(h_{j,m}+\boldsymbol{\rm h}_{{\rm r}, m}^H\boldsymbol{\rm \Theta}_{k,l}\boldsymbol{\rm g}_{{\rm r},j})\sqrt{p_{j,k,l}}x_{j,k,l}}_{\textrm{Uplink co-channel signal}}+n_{m,k,l},
\end{array}
\end{equation}
where $\boldsymbol{\rm v}_{k,l}\in \mathbb{C}^{M_{\rm T}\times 1}$ denotes the energy beamforming vector in phase $l$ of time slot $k$ with covariance matrix $\boldsymbol{\rm V}_{k,l}=\boldsymbol{\rm v}_{k,l}\boldsymbol{\rm v}_{k,l}^H$ and $\boldsymbol{\rm V}_{k,l}\succeq \boldsymbol{0}$. $\boldsymbol{\rm h}_{{\rm d}, m}\in\mathbb{C}^{M_{\rm T}\times 1}$ and $\boldsymbol{\rm h}_{{\rm r}, m}\in\mathbb{C}^{N\times 1}$ denote the channel vectors from the HAP and the RIS to the $m$-th IoT device, respectively. $h_{j,m}$ denotes the channel coefficient from the $j$-th IoT device to IoT device $m$. $\boldsymbol{\rm g}_{{\rm r}, j}\in \mathbb{C}^{N\times 1}$ is the counterpart uplink channel. $\boldsymbol{\rm H}\in\mathbb{C}^{N\times M_{\rm T}}$ denotes the channel matrix from the HAP to the RIS. $\boldsymbol{\rm \Theta}_{k,l} \triangleq {\rm diag}(e^{j\theta_{k,l,1}},\cdot\cdot\cdot,e^{j\theta_{k,l,N}})$ is the diagonal phase-shift matrix in phase $l$ of time slot $k$ with $\theta_{k,l,n}\in [0,2\pi]$ being the corresponding phase shift. $s_{k,l}$ and $x_{j,k,l}$ are the energy signal  transmitted by the HAP in phase $l$ of the $k$-th time slot and the transmit signal of IoT device $j$ in the group $G_{\hat l}$, respectively. $\mathbb{E}\{|s_{k,l}|^2\}=\mathbb{E}\{|x_{j,k,l}|^2\}=1$, $\hat{l} = 3-l$ that is complement to $l$. $p_{j,k,l}$ denotes the transmit power of IoT device $j$ in phase $l$ of the $k$-th time slot. $n_{m,k,l}$ is the additive white Gaussian noise at IoT device $m$ in phase $l$ of the $k$-th time slot.

To avoid the resource allocation mismatches with the traditional linear energy harvesting model, we adopt a more general and practical piece-wise linear energy harvesting model\cite{i0}, i.e.,
\begin{equation}\label{eq2}
\begin{array}{l}
E=
\left\{
\begin{array}{l}
\zeta P_{\rm EH},~~ \zeta P_{\rm EH}<P_{\rm sat},\\
P_{\rm sat},~~~~{\textrm{otherwise}},
\end{array}
\right.
\end{array}
\end{equation}
where $\zeta$ is the energy harvesting efficiency in the linear regime, $P_{\rm EH}$ is the received power, and $P_{\rm sat}$ is saturation power, beyond the saturation power, the received power will not increase. Based on the piece-wise energy harvesting model, the harvested energy of the $m$-th IoT device in phase $l$ of the $k$-th time slot will be
\begin{equation}\label{eq3}
\begin{array}{l}
E_{m,k,l}^{\rm IoT}=t_{k,l}\min\{\zeta P_{m,k,l}^{\rm IoT},P_{{\rm sat},m}^{\rm IoT}\},
\end{array}
\end{equation}
where $P_{{\rm sat},m}^{\rm IoT}$ denotes the saturation power of the $m$-th IoT device. $t_{k,l}$ denotes the $l$-th phase of the $k$-th time slot. Furthermore, the received power $P_{{m,k,l}}^{\rm IoT}$ is given by
\begin{equation}\label{eq4}
\begin{array}{l}
P_{{m,k,l}}^{\rm IoT}=|(\boldsymbol{\rm h}_{{\rm d},m}^H+\boldsymbol{\rm h}_{{\rm r}, m}^H\boldsymbol{\rm \Theta}_{k,l}\boldsymbol{\rm H})\boldsymbol{\rm v}_{k,l}|^2\\
~~~~~~~~+\sum\limits_{j\in G_{\hat{l}}}p_{j,k,l}|h_{j,m}+\boldsymbol{\rm h}_{{\rm r},m}^H\boldsymbol{\rm \Theta}_{k,l}\boldsymbol{\rm g}_{{\rm r},j}|^2.
\end{array}
\end{equation}

\subsection{The $\hat l$-th phase}
During the phase $\hat l$ of time slot $k$, the received signal at the HAP is written as
\begin{equation}\label{eq5}
\begin{array}{l}
\boldsymbol{\rm y}_{k,\hat l}^{\rm UL}=\underbrace{\sum\limits_{m\in G_{l}}(\boldsymbol{\rm g}_{{\rm d}, m}+\boldsymbol{\rm G}^H\boldsymbol{\rm \Theta}_{k,\hat l}\boldsymbol{\rm g}_{{\rm r}, m})\sqrt{p_{m,k,\hat l}}x_{m,k,\hat l}}_{\textrm{Desire signal}}\\
~~~~~+\underbrace{(\boldsymbol{\rm F}+\boldsymbol{\rm G}^H\boldsymbol{\rm \Theta}_{k,\hat l}\boldsymbol{\rm H})\boldsymbol{\rm v}_{k,\hat l}s_{k,\hat l}}_{\textrm{Self-interference}}+\boldsymbol{\rm n}_{k,\hat l},
\end{array}
\end{equation}
where $\boldsymbol{\rm g}_{{\rm d}, m}\in \mathbb{C}^{M_{\rm R}\times 1}$, $\boldsymbol{\rm G}\in \mathbb{C}^{ N\times M_{\rm R}}$, and $\boldsymbol{\rm g}_{{\rm r}, m}\in \mathbb{C}^{N\times 1}$ are the counterpart uplink channels. $\boldsymbol{\rm F}\in \mathbb{C}^{M_{\rm R}\times M_{\rm T}}$ represents the effective loopback channel at the HAP that satisfies $\mathbb{E}\{\|\boldsymbol{\rm F}\|_{\rm F}^2\}=\gamma$. $\boldsymbol{\rm n}_{k,\hat l}\in\mathcal{CN}(\boldsymbol{0},\delta^2\boldsymbol{\rm I}_{M_{\rm R}})$ denotes the additive white Gaussian noise at the HAP, and $\delta^2$ denotes the noise power.

To reduce the double propagation loss of the HAP-RIS-device link and interference introduced by the HAP-RIS-HAP link, we assume the RIS is deployed near IoT devices, away from the HAP. Thus, the self-interference term $\boldsymbol{\rm G}^H\boldsymbol{\rm \Theta}_{k,\hat l}\boldsymbol{\rm H}\boldsymbol{\rm v}_{k,\hat l}s_{k,\hat l}$ can be safely neglected \cite{g2}. Meanwhile, we consider an imperfect self-interference cancellation (SIC) process, this quantization error after analog-to-digital conversion (ADC) can be modeled as an independent white Gaussian noise \cite{c1}, i.e., $\boldsymbol{\rm n}_{{\rm error},k,\hat l}\sim\mathcal{CN}(\boldsymbol{0}, \beta\sigma_{{\rm error},k,\hat l}^2\boldsymbol{\rm I}_{M_{\rm R}})$, where $\beta\ll 1$ and $\sigma_{{\rm error},k,\hat l}^2$ is given by
\begin{equation}\label{eq6}
\begin{array}{l}
\sigma_{{\rm error},k,\hat l}^2=\|(\boldsymbol{\rm F}+\boldsymbol{\rm G}^H\boldsymbol{\rm \Theta}_{k,\hat l}\boldsymbol{\rm H})\boldsymbol{\rm v}_{k,\hat l}s_{k,\hat l}\|^2\approx \gamma\|\boldsymbol{\rm v}_{k,\hat l}\|^2.
\end{array}
\end{equation}

Therefore, the received signal at the HAP can be recast as
\begin{equation}\label{eq7}  
\begin{array}{l}
\boldsymbol{\rm \bar y}_{k,\hat l}^{\rm UL}=\sum\limits_{m\in G_{l}}(\boldsymbol{\rm g}_{{\rm d}, m}+\boldsymbol{\rm G}^H\boldsymbol{\rm \Theta}_{k,\hat l}\boldsymbol{\rm g}_{{\rm r}, m})\sqrt{p_{m,k,\hat l}}x_{m,k,\hat l}\\
~~~~~~+\boldsymbol{\rm n}_{{\rm error},k,\hat l}+\boldsymbol{\rm n}_{k,\hat l}.
\end{array}
\end{equation}

To decode the signal from IoT device $m$, the HAP applies a receive beamformer $\boldsymbol{\rm w}_{m,k,\hat l}$ to equalize the received signal for IoT device $m$ in phase $\hat l$ of the $k$-th time slot, such as
\begin{equation}\label{eq8}
\begin{array}{l}
\hat y_{m,k,\hat l}^{\rm UL}=\boldsymbol{\rm w}_{m,k,\hat l}^H\sum\limits_{m\in G_{l}}(\boldsymbol{\rm g}_{{\rm d}, m}+\boldsymbol{\rm G}^H\boldsymbol{\rm \Theta}_{k,\hat l}\boldsymbol{\rm g}_{{\rm r}, m})\sqrt{p_{m,k,\hat l}}x_{m,k,\hat l}\\
~~~~~~~+\boldsymbol{\rm w}_{m,k,\hat l}^H\boldsymbol{\rm n}_{{\rm error},k,\hat l}+\boldsymbol{\rm w}_{m,k,\hat l}^H\boldsymbol{\rm n}_{k,\hat l}.
\end{array}
\end{equation}

Then, the signal-to-interference-plus-noise ratio (SINR) of  the $m$-th IoT device’s signal recovered is given by
\begin{equation}\label{eq9}
\begin{array}{l}
\gamma_{m,k,\hat l}=\\
\frac{p_{m,k,\hat l}|\boldsymbol{\rm w}_{m,k,\hat l}^H(\boldsymbol{\rm g}_{{\rm d}, m}+\boldsymbol{\rm G}^H\boldsymbol{\rm \Theta}_{k,\hat l}\boldsymbol{\rm g}_{{\rm r}, m})|^2}{\sum\limits_{i\in G_{ l}\backslash \{m\}}p_{i,k,\hat l}|\boldsymbol{\rm w}_{m,k,\hat l}^H(\boldsymbol{\rm g}_{{\rm d}, i}+\boldsymbol{\rm G}^H\boldsymbol{\rm \Theta}_{k,\hat l}\boldsymbol{\rm g}_{{\rm r}, i})|^2+\beta\gamma\|\boldsymbol{\rm v}_{k,\hat l}\|^2+\delta^2}.
\end{array}
\end{equation}

Accordingly, the achievable throughput of the $m$-th IoT device during the $k$-th time slot is given by
\begin{equation}\label{eq10}
\begin{array}{l}
R_{k,m}=\sum\limits_{\hat l=1}^2\alpha_{m,l}t_{k,\hat l}\log_2(1+\gamma_{m,k,\hat l}),
\end{array}
\end{equation}
where $\alpha_{m, l}$ is a binary variable indicating the association of IoT device $m$ to a particular IoT group, $G_{l}$. For example, if $\alpha_{m,l}=1$, then IoT device $m$ undergoes uplink WIT in phase $\hat l$ and belongs to group $G_{l}$.

\subsection{Problem Formulation}
In this subsection, we formulate an optimization problem that maximizes the total throughput among all IoT devices. Mathematically, the problem is expressed as follows
\begin{equation}\label{eq11}
\begin{split}
&\max\limits_{\mbox{\scriptsize$\begin{array}{c} 
		\boldsymbol{\rm v}_{k,l},\boldsymbol{\rm \Theta}_{k,l},p_{m,k,\hat l},\\
		t_{k,l},\boldsymbol{\rm w}_{m,k,\hat l},\alpha_{m,l},
		\end{array}$}} 
\sum\limits_{k=1}^K\sum\limits_{m=1}^MR_{k,m}\\
s.t.~&{C_1}:~R_{k,m}\geq R_{k,m}^{\min},\\
&{C_2}:~t_{k,\hat l}p_{m,k,\hat l}+t_{k,l}p_{{\rm c},m}^{\rm passive}+t_{k,\hat l}p_{{\rm c},m}^{\rm active}\leq E_{m,k,l}^{\rm IoT},\\
&{C_3}:~\|\boldsymbol{\rm v}_{k,l}\|^2\leq P^{\max},\\
\end{split}
\end{equation}
\begin{equation}\nonumber 
\begin{split}
&{C_4}:~\left|[\boldsymbol{\rm \Theta}_{k,l}]_{n,n}\right|=1,\\
&{C_5}:~\alpha_{m,l}\in\{0,1\},\sum\limits_{ l=1}^2\alpha_{m,l}=1,\\
&{C_6}:~\sum\limits_{k=1}^K(t_{k,l}+t_{k,\hat l})= T, t_{k,l},t_{k,\hat l}>0,\\
&{C_7}:~\|\boldsymbol{\rm w}_{m,k,\hat l}\|^2=1,\\
\end{split}
\end{equation}
where $C_1$ denotes the minimum throughput constraint of each IoT device, and $R_{k,m}^{\min}$ is the minimum throughput threshold of IoT device $m$; $C_2$ denotes that the total energy consumption of the $m$-th IoT device should not exceed the total harvested energy, $p_{{\rm c},m}^{\rm passive}$ and $p_{{\rm c},m}^{\rm active}$ denote the circuit power consumption during the WET and WIT phases, respectively; $C_3$ denotes the maximum transmit power constraint of the HAP, and $P^{\max}$ denotes the maximum transmit power threshold of the HAP; $C_4$ guarantees that the diagonal phase-shift matrix $\boldsymbol{\rm \Theta}_{k,l}$ has $N$ unit modulus components on its main diagonal; $C_5$ indicates that each IoT device can be assigned to one group; $C_6$ denotes the total time constraint; $C_7$ denotes the receive beamforming constraint. We note that problem (\ref{eq11}) is a highly non-convex optimization problem. In particular, the coupling of the optimization variables, the non-convexity of the objective function, the unit-modulus constraint, and the discrete constraint are the main obstacles to solving the considered resource allocation problem efficiently. Thus, the globally optimal solution to this problem is in general intractable. In the next section, we develop a sub-optimal AO-based iterative algorithm to solve problem (\ref{eq11}) with a polynomial time complexity.
\section{The Algorithm for Group Switching-based Protocol}
For any given $\boldsymbol{{\rm \Theta}}_{k,l}$, $\boldsymbol{{\rm v}}_{k,\hat l}$, $p_{m,k,\hat l}$, $t_{k,l}$, and $\alpha_{m,l}$, it is well-known that the linear MMSE detector is the optimal receive beamforming to problem (\ref{eq11}). The MMSE-based receive beamforming is written as
\begin{equation}\label{eq12}
\begin{array}{l}
\boldsymbol{{\rm\bar w}}_{m,k,\hat l}^*=\{\sum\limits_{m\in G_{l}}p_{m,k,\hat l}\boldsymbol{\rm g}_{m,k,\hat l}\boldsymbol{\rm g}_{m,k,\hat l}^H+\beta\gamma\|\boldsymbol{\rm v}_{k,\hat l}\|^2\boldsymbol{\rm I}_{M_{\rm R}}\\
~~~~~~~~+\delta^2\boldsymbol{\rm I}_{M_{\rm R}}\}^{-1}\sqrt{p_{m,k,\hat l}}\boldsymbol{\rm g}_{m,k,\hat l},
\end{array}
\end{equation}
where $\boldsymbol{\rm g}_{m,k,\hat l}=\boldsymbol{\rm g}_{{\rm d}, m}+\boldsymbol{\rm G}^H\boldsymbol{\rm \Theta}_{k,\hat l}\boldsymbol{\rm g}_{{\rm r}, m}$. Then, we have $\boldsymbol{{\rm w}}_{m,k,\hat l}^*=\frac{\boldsymbol{{\rm\bar w}}_{m,k,\hat l}^*}{\|\boldsymbol{{\rm\bar w}}_{m,k,\hat l}^*\|^2}$. With $\boldsymbol{{\rm w}}_{m,k,\hat l}^*$, the SINR of the $m$-th IoT device in phase $\hat l$ of time slot $k$ is shown in (\ref{eq13}) at the top of next page.
\begin{figure*}
\begin{equation}\label{eq13}
\begin{array}{l}
\bar\gamma_{m,k,\hat l}=
\dfrac{1}{\dfrac{1}{p_{m,k,\hat l}\boldsymbol{\rm g}_{m,k,\hat l}^H\{\sum\limits_{m\in G_{l}}p_{m,k,\hat l}\boldsymbol{\rm g}_{m,k,\hat l}\boldsymbol{\rm g}_{m,k,\hat l}^H+\beta\gamma\|\boldsymbol{\rm v}_{k,\hat l}\|^2\boldsymbol{\rm I}_{M_{\rm R}}+\delta^2\boldsymbol{\rm I}_{M_{\rm R}}\}^{-1}\boldsymbol{\rm g}_{m,k,\hat l}}-1}.
\end{array}
\end{equation}	
\hrulefill
\end{figure*}


Then, the optimization problem (\ref{eq11}) can be reformulated as
\begin{equation}\label{eq14}
\begin{array}{l}
\max\limits_{\mbox{\scriptsize$\begin{array}{c} 
		\boldsymbol{\rm \Theta}_{k,l},p_{m,k,\hat l},\\
		\boldsymbol{\rm v}_{k,\hat l},
		t_{k,l},\alpha_{m,l},
		\end{array}$}} 
\sum\limits_{k=1}^K\sum\limits_{m=1}^MR_{k,m}\\
~~~~~~~~~~~~~s.t.~{C_1}-{C_6}.
\end{array}
\end{equation}
It can be seen that the coupled relationship among optimization variables poses a challenge in solving problem (\ref{eq14}). To this end, we develop an AO method to solve this problem in an iterative manner.
\subsection{Optimizing energy beamforming $\boldsymbol{\rm v}_{k,\hat l}$ and phase shift $\boldsymbol{\rm \Theta}_{k,l}$}
For any given $\boldsymbol{{\rm w}}_{m,k,\hat l}$, $p_{m,k,\hat l}$, $t_{k,l}$, and $\alpha_{m,l}$, problem (\ref{eq14}) can be reformulate as
\begin{equation}\label{eq15}
\begin{array}{l}
\max\limits_{\mbox{\scriptsize$\begin{array}{c} 
		\boldsymbol{\rm \Theta}_{k,l},\boldsymbol{\rm v}_{k,\hat l}
		
		\end{array}$}} 
\sum\limits_{k=1}^K\sum\limits_{m=1}^MR_{k,m}\\
~~~~~~~~~s.t.~{C_1}-{C_4}.
\end{array}
\end{equation}
The convex transformation of $R_{k,m}$ is challenging due to the fact that both the numerator and denominator in $R_{k,m}$ contain optimization variables. Based on the SCA method, we substitute the numerators and the
denominators of $R_{k,m}$ by slack variables, i.e.,
\begin{equation}\label{eq16}
\begin{array}{l}
C_{1-1}:~\sum\limits_{m=1}^Mp_{m,k,\hat l}|\boldsymbol{\rm w}_{m,k,\hat l}^H(\boldsymbol{\rm g}_{{\rm d}, m}+\boldsymbol{\rm G}^H\boldsymbol{\rm \Theta}_{k,\hat l}\boldsymbol{\rm g}_{{\rm r}, m})|^2\\
~~~~~~~~~~+\beta\gamma\|\boldsymbol{\rm v}_{k,\hat l}\|^2+\delta^2\geq e^{u_{m,k,\hat l}},
\end{array}
\end{equation}
\begin{equation}\label{eq17}
\begin{array}{l}
C_{1-2}:~\sum\limits_{i\in G_{ l}\backslash \{m\}}p_{i,k,\hat l}|\boldsymbol{\rm w}_{m,k,\hat l}^H(\boldsymbol{\rm g}_{{\rm d}, i}{+}\boldsymbol{\rm G}^H\boldsymbol{\rm \Theta}_{k,\hat l}\boldsymbol{\rm g}_{{\rm r}, i})|^2\\
~~~~~~~~~~+\beta\gamma\|\boldsymbol{\rm v}_{k,\hat l}\|^2+\delta^2\leq e^{v_{m,k,\hat l}}.
\end{array}
\end{equation}

To facilitate the design of an efficient algorithm, the term $|\boldsymbol{\rm w}_{m,k,\hat l}^H(\boldsymbol{\rm g}_{{\rm d}, m}+\boldsymbol{\rm G}^H\boldsymbol{\rm \Theta}_{k,\hat l}\boldsymbol{\rm g}_{{\rm r}, m})|^2$ can be transformed into
\begin{equation}\label{eq18}
\begin{array}{l}
|\boldsymbol{\rm w}_{m,k,\hat l}^H(\boldsymbol{\rm g}_{{\rm d}, m}{+}\boldsymbol{\rm G}^H\boldsymbol{\rm \Theta}_{k,\hat l}\boldsymbol{\rm g}_{{\rm r}, m})|^2{=}{\rm Tr}(\boldsymbol{{\rm G}}_m^H\boldsymbol{{\rm O}}_{k,\hat l}\boldsymbol{{\rm G}}_m\boldsymbol{\rm W}_{m,k,\hat l}),
\end{array}
\end{equation}
where $\boldsymbol{{\rm o}}_{k,l}=[\boldsymbol{{\rm \theta}}_{k,l};1]$ and $\boldsymbol{{\rm O}}_{k,l}=\boldsymbol{{\rm o}}_{k,l}\boldsymbol{{\rm o}}_{k,l}^H$, when ${\rm Rank}(\boldsymbol{{\rm O}}_{k,l})=1$ holds, $\boldsymbol{{\rm \Theta}}_{k,l}$ can be denoted by ${\rm diag}([\boldsymbol{{\rm O}}_{k,l}]_{N+1,1:N})$, where $[\boldsymbol{{\rm O}}_{k,l}]_{N+1,1:N}=\left[[\boldsymbol{{\rm O}}_{k,l}]_{N+1,1},\cdot\cdot\cdot,[\boldsymbol{{\rm O}}_{k,l}]_{N+1,N}\right]$. $\boldsymbol{{\rm G}}_m{=}
[{\rm diag}(\boldsymbol{{\rm g}}_{{\rm r},m}^H)\boldsymbol{{\rm G}};
\boldsymbol{{\rm g}}_{{\rm d},m}^H]$ and $\boldsymbol{\rm W}_{m,k,\hat l}=\boldsymbol{\rm w}_{m,k,\hat l}\boldsymbol{\rm w}_{m,k,\hat l}^H$. Thus, $C_{1-1}$ can be transformed into
\begin{equation}\label{eq19}
\begin{array}{l}
\bar C_{1-1}:\sum\limits_{m=1}^Mp_{m,k,\hat l}{\rm Tr}(\boldsymbol{{\rm G}}_m^H\boldsymbol{{\rm O}}_{k,\hat l}\boldsymbol{{\rm G}}_m\boldsymbol{\rm W}_{m,k,\hat l})\\
~~~~~~~~~+\beta\gamma\|\boldsymbol{\rm v}_{k,\hat l}\|^2+\delta^2\geq e^{u_{m,k,\hat l}}.
\end{array}
\end{equation}

For $C_{1-2}$, we apply the first order Taylor expansion to obtain its upper bound, i.e.,
\begin{equation}\label{eq20}
\begin{array}{l}
\bar C_{1-2}:~\sum\limits_{i\in G_{ l}\backslash \{m\}}p_{i,k,\hat l}{\rm Tr}(\boldsymbol{{\rm G}}_i^H\boldsymbol{{\rm O}}_{k,\hat l}\boldsymbol{{\rm G}}_i\boldsymbol{\rm W}_{m,k,\hat l}){+}\beta\gamma\|\boldsymbol{\rm v}_{k,\hat l}\|^2\\
~~~~~~~~~~~~+\delta^2\leq e^{\bar v_{m,k,\hat l}}(v_{m,k,\hat l}-\bar v_{m,k,\hat l}).
\end{array}
\end{equation}

Next, we deal with $E_{m,k,l}^{\rm IoT}$ in $C_2$, i.e.,
\begin{equation}\label{eq21}
\begin{array}{l}
E_{m,k,l}^{\rm IoT}{=} t_{k,l}\min\{\zeta(|(\boldsymbol{\rm h}_{{\rm d}, m}^H+\boldsymbol{\rm h}_{{\rm r}, m}^H\boldsymbol{\rm \Theta}_{k,l}\boldsymbol{\rm H})\boldsymbol{\rm v}_{k,l}|^2\\
~~~~~~~{+}\sum\limits_{j\in G_{\hat l}} p_{j,k,l}|h_{j,m}{+}\boldsymbol{\rm h}_{{\rm r},m}^H\boldsymbol{\rm \Theta}_{k,l}\boldsymbol{\rm g}_{{\rm r},j}|^2),P_{{\rm sat},m,k,l}^{\rm IoT}\} \\
{=} t_{k,l}\min\{\zeta({\rm Tr}(\boldsymbol{\rm H}_m\boldsymbol{\rm V}_{k,l}\boldsymbol{\rm H}_m^H\boldsymbol{\rm O}_{k,l})\\
~~~~~~~{+}\sum\limits_{j\in G_{\hat l}} p_{j,k,l}{\rm Tr}(\boldsymbol{\rm H}_{j,m}\boldsymbol{\rm O}_{k,l})),P_{{\rm sat},m}^{\rm IoT}\},\\
\end{array}
\end{equation}
where $\boldsymbol{\rm V}_{k,l}=\boldsymbol{\rm v}_{k,l}\boldsymbol{\rm v}_{k,l}^H$ with $\boldsymbol{\rm V}_{k,l}\succeq\boldsymbol{{0}}$, given $\boldsymbol{\rm V}_{k,l}$, $\boldsymbol{\rm v}_{k,l}$ 
can be recovered from the eigenvalue decomposition. $\boldsymbol{{\rm H}}_{m}=
[{\rm diag}(\boldsymbol{{\rm h}}_{{\rm r},m}^H)\boldsymbol{{\rm H}};
\boldsymbol{{\rm h}}_{{\rm d},m}^H]$, $\boldsymbol{{\rm h}}_{j,m}=
[{\rm diag}(\boldsymbol{{\rm h}}_{{\rm r},m}^H)\boldsymbol{{\rm g}}_{{\rm r},j};
h_{j,m}]
$, $\boldsymbol{{\rm H}}_{j,m}=\boldsymbol{{\rm h}}_{j,m}\boldsymbol{{\rm h}}_{j,m}^H$.


There are some challenges in solving for ${\rm Tr}(\boldsymbol{\rm H}_m\boldsymbol{\rm V}_{k,l}\boldsymbol{\rm H}_m^H\boldsymbol{\rm O}_{k,l})$, unlike most of the existing works adopting the AO method which optimizes $\boldsymbol{{\rm V}}_{k,l}$ and $\boldsymbol{\rm O}_{k,l}$ separately, we aim to jointly optimize $\boldsymbol{{\rm V}}_{k,l}$ and $\boldsymbol{\rm O}_{k,l}$. However, the multiplication of two matrices poses a challenge in solving our problem. To this end, we further rewrite the related terms as
\begin{equation}\label{eq22}
\begin{array}{l}
{\rm Tr}(\boldsymbol{\rm H}_m\boldsymbol{\rm V}_{k,l}\boldsymbol{\rm H}_m^H\boldsymbol{\rm O}_{k,l})=\frac{1}{2}\|\boldsymbol{\rm O}_{k, l}+\boldsymbol{{\rm H}}_m\boldsymbol{\rm V}_{k, l}\boldsymbol{{\rm H}}_m^H\|_{\rm F}^2\\
~~~~~~~~~-\frac{1}{2}\|\boldsymbol{\rm O}_{k, l}\|_{\rm F}^2-\frac{1}{2}\|\boldsymbol{{\rm H}}_m\boldsymbol{\rm V}_{k, l}\boldsymbol{{\rm H}}_m^H\|_{\rm F}^2.
\end{array}
\end{equation}
However, the first term of (\ref{eq22}) is not concave. To handle it, we establish the corresponding lower bounds via the first-order Taylor approximation. The term $\frac{1}{2}\|\boldsymbol{\rm O}_{k, l}+\boldsymbol{{\rm H}}_m\boldsymbol{\rm V}_{k, l}\boldsymbol{{\rm H}}_m^
H\|_{\rm F}^2$ can be bounded by an affine function which is given by
\begin{equation}\label{eq23}
\begin{array}{l}
\frac{1}{2}\|\boldsymbol{\rm O}_{k, l}+\boldsymbol{{\rm H}}_m\boldsymbol{\rm V}_{k, l}\boldsymbol{{\rm H}}_m^
H\|_{\rm F}^2\geq \frac{1}{2}\|\boldsymbol{\rm \bar O}_{k, l}+\boldsymbol{{\rm H}}_m\boldsymbol{\rm\bar V}_{k, l}\boldsymbol{{\rm H}}_m^H\|_{\rm F}^2\\
+{\rm Tr}((\boldsymbol{\rm \bar O}_{k, l}+\boldsymbol{{\rm H}}_m\boldsymbol{\rm\bar V}_{k, l}\boldsymbol{{\rm H}}_m^
H)^H(\boldsymbol{\rm O}_{k, l}-\boldsymbol{\rm \bar O}_{k, l}))\\
+{\rm Tr}((\boldsymbol{\rm H}_m^H\boldsymbol{\rm \bar O}_{k, l}\boldsymbol{\rm H}_m
{+}\boldsymbol{\rm H}_m^H\boldsymbol{\rm H}_m\boldsymbol{\rm\bar V}_{k, l}\boldsymbol{\rm H}_m^H\boldsymbol{\rm H}_m)^H(\boldsymbol{\rm V}_{k, l}{-}\boldsymbol{\rm\bar V}_{k, l})).
\end{array}
\end{equation}

Then, $\bar C_2$ can be rewritten as
\begin{equation}\label{eq24}
\begin{array}{l}
{\bar C_2}:t_{k,\hat l}p_{m,k,\hat l}{+}p_{{\rm c},m}^{\rm passive}t_{k,l}{+}p_{{\rm c},m}^{\rm active}t_{k,\hat l}\leq  \bar E_{m,k,l}^{\rm IoT},
\end{array}
\end{equation}
where
\begin{equation}\label{eq25}
\begin{array}{l}
\bar E_{m,k,l}^{\rm IoT}=t_{k,l}\min\{\zeta(\frac{1}{2}\|\boldsymbol{\rm \bar O}_{k, l}+\boldsymbol{{\rm H}}_m\boldsymbol{\rm\bar V}_{k, l}\boldsymbol{{\rm H}}_m^
H\|_{\rm F}^2\\
+{\rm Tr}((\boldsymbol{\rm \bar O}_{k, l}+\boldsymbol{{\rm H}}_m\boldsymbol{\rm\bar V}_{k, l}\boldsymbol{{\rm H}}_m^
H)^H(\boldsymbol{\rm O}_{k, l}-\boldsymbol{\rm \bar O}_{k, l}))\\
+{\rm Tr}((\boldsymbol{\rm H}_m^H\boldsymbol{\rm \bar O}_{k, l}\boldsymbol{\rm H}_m
{+}\boldsymbol{\rm H}_m^H\boldsymbol{\rm H}_m\boldsymbol{\rm\bar V}_{k, l}\boldsymbol{\rm H}_m^H\boldsymbol{\rm H}_m)^H(\boldsymbol{\rm V}_{k, l}{-}\boldsymbol{\rm\bar V}_{k, l}))\\
-\frac{1}{2}\|\boldsymbol{\rm O}_{k, l}\|_{\rm F}^2
-\frac{1}{2}\|\boldsymbol{{\rm H}}_m\boldsymbol{\rm V}_{k, l}\boldsymbol{{\rm H}}_m^H\|_{\rm F}^2\\
+\sum\limits_{j\in G_{\hat l}} p_{j,k,l}{\rm Tr}(\boldsymbol{\rm H}_{j,m}\boldsymbol{\rm O}_{k,l})),P_{{\rm sat},m}^{\rm IoT}\}.
\end{array}
\end{equation}

Accordingly, problem (\ref{eq15}) is transformed as the following optimization problem
\begin{equation}\label{eq26}
\begin{array}{l}
\max\limits_{\mbox{\scriptsize$\begin{array}{c} 
		\boldsymbol{\rm V}_{k,l},\boldsymbol{\rm O}_{k,l},\\
		u_{m,k,\hat l},v_{m,k,\hat l}
		\end{array}$}} 
\sum\limits_{k=1}^K\sum\limits_{m=1}^M\sum\limits_{\hat l=1}^2\frac{1}{\ln2}\alpha_{m,l}t_{k,\hat l}(u_{m,k,\hat l}-v_{m,k,\hat l})\\
s.t.~{\bar C_{1-1}},{\bar C_{1-2}},{\bar C_2},\\
~~~~~{\bar C_1}:~\sum\limits_{\hat l=1}^2\frac{1}{\ln2}\alpha_{m,l}t_{k,\hat l}(u_{m,k,\hat l}-v_{m,k,\hat l})\geq R_{k,m}^{\min},\\
~~~~~\bar {C_3}:~{\rm Tr}(\boldsymbol{\rm V}_{k,l})\leq P^{\max},\boldsymbol{{\rm V}}_{m,k,\hat l}\succeq\boldsymbol{{0}},\\
~~~~~{\bar C_4}:~[\boldsymbol{\rm O}_{k,l}]_{n,n}\leq1,\boldsymbol{\rm O}_{k,l}\succeq\boldsymbol{{0}},[\boldsymbol{\rm O}_{k,l}]_{N+1,N+1}=1,\\
~~~~~{C_8}:~{\rm Rank}(\boldsymbol{\rm O}_{k,l})=1.\\
\end{array}
\end{equation}
It is worth noting that the rank-one constraint $C_8$ is an obstacle to solving problem (\ref{eq26}). Most existing works adopt the semi-definite relaxation method to deal with the rank-one constraint. However, the semi-definite relaxation method may not result in a rank-one matrix $\boldsymbol{{\rm O}}_{k,l}$. Moreover, 
some approximation methods such as the Gaussian randomization method may deteriorate system performance when the number of antennas becomes large. To tackle this obstacle, we first equivalently transform $C_8$ into the following difference of convex functions constraint\cite{i1}
\begin{equation}\label{eq27}
\begin{array}{l}
{\bar C_8}:~\|\boldsymbol{\rm O}_{k,l}\|_*-\|\boldsymbol{\rm O}_{k,l}\|_2=0,\\
\end{array}
\end{equation}
where the inequality  $\|\boldsymbol{\rm O}_{k,l}\|_*=\sum\limits_{p}\sigma_p(\boldsymbol{\rm O}_{k,l})\geq\|\boldsymbol{\rm O}_{k,l}\|_2=\max\limits_{p}\sigma_p(\boldsymbol{\rm O}_{k,l})$ and the equality holds if and only if $\boldsymbol{\rm O}_{k,l}$ is a rank-one matrix, $\sigma_p(\boldsymbol{\rm O}_{k,l})$ denotes the $p$-th singular value of $\boldsymbol{\rm O}_{k,l}$. Then, we deal with the rank-one constraint $C_8$ by applying a penalty-based method. Thus, problem (\ref{eq26}) is transformed to
\begin{equation}\label{eq28}
\begin{array}{l}
\max\limits_{\mbox{\scriptsize$\begin{array}{c} 
		\boldsymbol{\rm V}_{k,l},\boldsymbol{\rm O}_{k,l},\\
		u_{m,k,\hat l},v_{m,k,\hat l}
		\end{array}$}} 
\sum\limits_{k=1}^K\sum\limits_{m=1}^M\sum\limits_{\hat l=1}^2\frac{1}{\ln2}\alpha_{m,l}t_{k,\hat l}(u_{m,k,\hat l}-v_{m,k,\hat l})\\
~~~~~~~~~~~~~~~~~~~~-\frac{1}{2\mu}(\|\boldsymbol{\rm O}_{k,l}\|_*-\|\boldsymbol{\rm O}_{k,l}\|_2)\\
~~~~~~~~~~~~~~~~s.t.~{\bar C_1}-{\bar C_4},{\bar C_{1-1}},{\bar C_{1-2}},\\
\end{array}
\end{equation}
where $\mu>0$ is the penalty factor which penalizes the objective function for any $\boldsymbol{\rm O}_{k,l}$ whose rank is larger than one.  By gradually decreasing the value of $\mu$, $\frac{1}{2\mu}\rightarrow \infty$, then  we can obtain a rank-one solution of problem (\ref{eq28}). Furthermore, $\|\boldsymbol{\rm O}_{k,l}\|_2$ can be linearized, i.e.,
\begin{equation}\label{eq29}
\begin{array}{l}
\|\boldsymbol{\rm O}_{k,l}\|_2{\geq} \|\boldsymbol{\rm\bar O}_{k,l}\|_2{+}{\rm Tr}(\boldsymbol{{\rm\lambda}}_{1,\max}\boldsymbol{{\rm\lambda}}_{1,\max}^H(\boldsymbol{\rm O}_{k,l}{-}\boldsymbol{\rm\bar O}_{k,l}))\\
~~~~~~~~~\triangleq { O}_{k,l},
\end{array}
\end{equation}
where $\boldsymbol{{\rm\lambda}}_{1,\max}$ is the eigenvector associated with the principal eigenvalue of $\boldsymbol{\rm\bar O}_{k,l}$. Therefore, the optimization problem (\ref{eq28}) is transformed into
\begin{equation}\label{eq30}
\begin{array}{l}
\max\limits_{\mbox{\scriptsize$\begin{array}{c} 
		\boldsymbol{\rm V}_{k,l},\boldsymbol{\rm O}_{k,l},\\
		u_{m,k,\hat l},v_{m,k,\hat l}
		\end{array}$}} 
\sum\limits_{k=1}^K\sum\limits_{m=1}^M\sum\limits_{\hat l=1}^2\frac{1}{\ln2}\alpha_{m,l}t_{k,\hat l}(u_{m,k,\hat l}-v_{m,k,\hat l})\\
~~~~~~~~~~~~~~~~~~~~-\frac{1}{\mu}(\|\boldsymbol{\rm O}_{k,l}\|_*-{ O}_{k,l})\\
~~~~~~~~~~~~s.t.~{\bar C_1}-{\bar C_4},{\bar C_{1-1}},{\bar C_{1-2}}.\\
\end{array}
\end{equation}
Problem (\ref{eq30}) is convex and thus can be solved by the standard convex optimization techniques.
\subsection{Optimizing uplink transmit power $p_{m,k,\hat l}$}
For given $\boldsymbol{{\rm v}}_{k,l}$, $\boldsymbol{{\rm w}}_{m,k,\hat l}$, $\boldsymbol{\rm \Theta}_{k,l}$, $\alpha_{m,l}$, and $t_{k,l}$, we can optimize $p_{m,k,\hat l}$. The optimization problem (\ref{eq14}) is reduced as the following sub-problem
\begin{equation}\label{eq31}
\begin{array}{l}
\max\limits_{\mbox{\scriptsize$\begin{array}{c} 
		p_{m,k,\hat l},\\
		u_{m,k,\hat l},
		v_{m,k,\hat l}
		\end{array}$}}
	 \sum\limits_{m=1}^M\sum\limits_{k=1}^K\sum\limits_{\hat l=1}^2\frac{1}{\ln2}\alpha_{m,l}t_{k,\hat l}(u_{m,k,\hat l}{-}v_{m,k,\hat l})\\
~~~~~~~~~~~~~~~~s.t.~{\bar C_{1}},{\bar C_{1-1}},{\bar C_{1-2}},{\bar C_2}.\\
\end{array}
\end{equation}
Problem (\ref{eq31}) is a convex optimization problem and can be solved directly by the standard convex optimization techniques, such as the interior-point method.
\begin{figure*}
	\centering
	\includegraphics[width=6.5in]{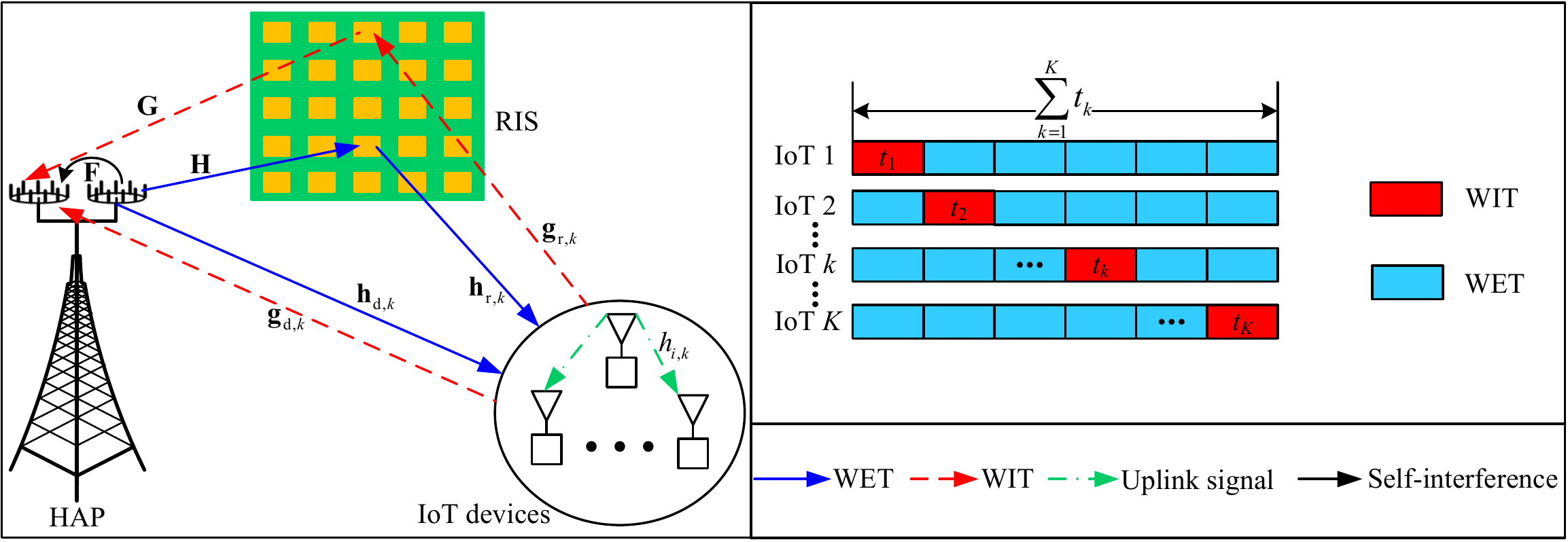}
	\caption{An RIS-aided WPCN with user switching.}
	\label{fig2}
\end{figure*}
\subsection{Optimizing the grouping factor $\alpha_{m,l}$ and transmission time $t_{k,l}$}
For given $p_{m,k,\hat l}$, $\boldsymbol{{\rm v}}_{k,l}$, $\boldsymbol{{\rm w}}_{m,k,\hat l}$, and $\boldsymbol{\rm \Theta}_{k,l}$, we further optimize $\alpha_{m,l}$ and $t_{k,l}$. The optimization problem (\ref{eq14}) is reduced as the following sub-problem 
\begin{equation}\label{eq32}
\begin{array}{l}
\max\limits_{\mbox{\scriptsize$\begin{array}{c} 
		t_{k,l},\alpha_{m,l},
		\end{array}$}} 
\sum\limits_{m=1}^M\sum\limits_{k=1}^K\sum\limits_{\hat l=1}^2\alpha_{m,l}t_{k,\hat l}\log_2(1+\bar\gamma_{m,k,\hat l})\\
s.t.~{C_2},{C_5},{C_6},\\
~~~~~{\tilde C_1}:~\sum\limits_{\hat l=1}^2\alpha_{m,l}t_{k,\hat l}\log_2(1+\bar\gamma_{m,k,\hat l})\geq R_{k,m}^{\min}.\\
\end{array}
\end{equation}
However, problem (\ref{eq32}) is challenging to solve due to $C_5$. To solve this problem efficiently, we first relax the grouping factor $\alpha_{m,l}$ to a continuous real variable in the range of $[0,1]$ by using convex relaxation \cite{i3}, thus we could consider it as a time-sharing factor. As a result,  problem (\ref{eq32}) can be rewritten as
\begin{equation}\label{eq33}
\begin{array}{l}
\max\limits_{\mbox{\scriptsize$\begin{array}{c} 
		t_{k,l},\alpha_{m,l},
		\end{array}$}} 
\sum\limits_{m=1}^M\sum\limits_{k=1}^K\sum\limits_{\hat l=1}^2\frac{1}{\ln2}\alpha_{m,l}t_{k,\hat l}\log_2(1+\bar\gamma_{m,k,\hat l})\\
~~~~~~s.t.~{\tilde C_1},{C_2},{C_6},{\bar C_5}:~\alpha_{m,l}\in [0,1].
\end{array}
\end{equation}
Furthermore, defining $\bar t_{m,k,\hat l,l}=\alpha_{m,l}t_{k,\hat l}$, the problem (\ref{eq33}) can be transformed as
\begin{equation}\label{eq34}
\begin{array}{l}
\max\limits_{\mbox{\scriptsize$\begin{array}{c} 
		t_{k,l},\bar t_{m,k,\hat l,l}
		\end{array}$}} 
\sum\limits_{m=1}^M\sum\limits_{k=1}^K\sum\limits_{\hat l=1}^2\frac{1}{\ln2}\bar t_{m,k,\hat l,l}\log_2(1+\gamma_{m,k,\hat l})\\
~~~~~~s.t.~{C_2},{C_6},{\hat C_1}:~\sum\limits_{\hat l=1}^2\bar t_{m,k,\hat l,l}\log_2(1{+}\gamma_{m,k,\hat l}){\geq} R_{k,m}^{\min},\\
~~~~~~~~~~~{\tilde C_5}:~0\leq \bar t_{m,k,\hat l,l}\leq t_{k,\hat l}.\\
\end{array}
\end{equation}
Problem (\ref{eq34}) is a convex optimization problem and can be solved directly by the standard convex optimization techniques. The group switching-based algorithm is shown in \textbf{Algorithm 1}.
\begin{spacing}{1.00}
	\floatname{algorithm}{Algorithm}
	\renewcommand{\algorithmicrequire}{\textbf{Input:}}
	\renewcommand{\algorithmicensure}{\textbf{Output:}}
	\begin{algorithm}[!t]
		\small
		\caption{The Group Switching-based Algorithm }
		\begin{algorithmic}[1]
			\State Initialize system parameters: $K$, $M$, $N$, $T$, $P^{\max}$, $R_{k,m}^{\min}$, $\delta^2$, $P_{{\rm sat},m,k,l}^{\rm IoT}$, $p_{{\rm c},m}^{\rm active}$, $p_{{\rm c},m}^{\rm passive}$, $\beta$, $\gamma$, $\zeta$, $\boldsymbol{\rm\bar V}_{k, l}$,$\boldsymbol{\rm \bar O}_{k, l}$, $u_{m,k,\hat l}$, $v_{m,k,\hat l}$;
			\State Set the maximum iteration number $L_{\max}$ and the convergence accuracy $\epsilon$, set the initial iteration index $ll=0$;
			\While{$ll\leq L_{\max}$}
			\State Initialize $\boldsymbol{{\rm \Theta}}_{k,l}$, $\alpha_{m,l}$, $p_{m,k,\hat l}$, $\boldsymbol{{\rm v}}_{k,\hat l}$, and $t_{k,l}$;
			\State Calculate $\boldsymbol{{\rm w}}_{m,k,\hat l}$ via (\ref{eq12});
			\Repeat
			\State Given $\boldsymbol{{\rm w}}_{m,k,\hat l}$, $\alpha_{m,l}$, $p_{m,k,\hat l}$, and $t_{k,l}$, Calculate $\boldsymbol{{\rm O}}_{k,l}$ and $\boldsymbol{{\rm V}}_{k,\hat l}$ via (\ref{eq30});
			\Until{converges};
			\State $\boldsymbol{{\rm \Theta}}_{k,l}={\rm diag}([\boldsymbol{{\rm O}}_{k,l}]_{N+1,1:N})$ and $\boldsymbol{{\rm v}}_{k,\hat l}$ can be obtained by applying the eigenvalue decomposition method.
			\Repeat
			\State Given $\boldsymbol{{\rm w}}_{m,k,\hat l}$, $\alpha_{m,l}$, $\boldsymbol{{\rm \Theta}}_{k,l}$, $\boldsymbol{{\rm v}}_{k,\hat l}$, and $t_{k,l}$, Calculate $p_{m,k,\hat l}$ via (\ref{eq31});
			\Until{converges};
			\State Given $\boldsymbol{{\rm w}}_{m,k,\hat l}$, $\boldsymbol{{\rm \Theta}}_{k,l}$, $\boldsymbol{{\rm v}}_{k,\hat l}$, and $p_{m,k,\hat l}$, Calculate $t_{k,l}$ and $\alpha_{m,l}$ via (\ref{eq34});
			\If{$|\sum\limits_{k=1}^K\sum\limits_{m=1}^MR_{k,m}(ll)-\sum\limits_{k=1}^K\sum\limits_{m=1}^MR_{k,m}(ll-1)|\leq\epsilon$}
			\State \textbf{break}.
			\Else
			\State $ll=ll+1$;
			\EndIf
			\EndWhile		
		\end{algorithmic}
	\end{algorithm}
\end{spacing} 
\section{The User Switching-based Protocol}
The user switching-based protocol is investigated in this section. Consider an RIS-aided WPCN consisting of an HAP, $K$ single-antenna IoT devices, and an RIS, as shown in Fig. \ref{fig2}. The time frame $T$ is divided into $K$ time slots, the HAP transmits wireless energy to IoT devices with the help of the RIS during the total frame. Meanwhile, each IoT device occupies one time slot to transmit wireless information to the HAP based on a TDMA manner.
\subsection{The WET Phase}
\begin{figure}
	\centering
	\includegraphics[width=3in]{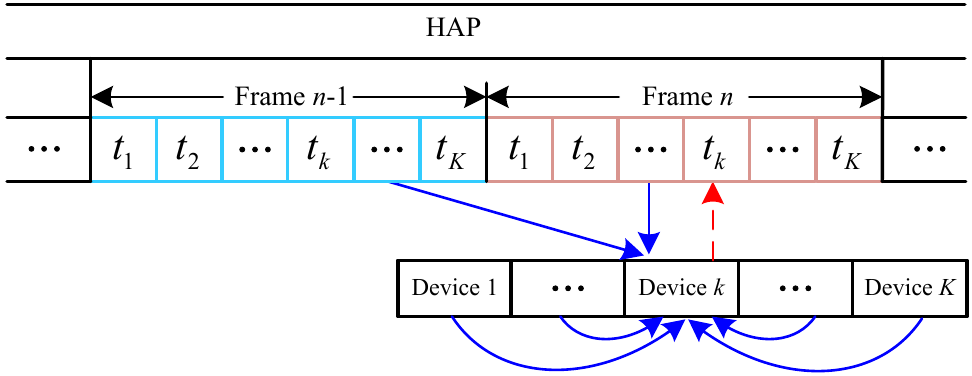}
	\caption{The periodic transmission protocol.}
	\label{fig3}
\end{figure}

In the downlink WET phase, the received signal of the $k$-th IoT device in time slot $i$ is given by

\begin{equation}\label{eq35}
\begin{array}{l}
y_{k,i}^{\rm DL}=\underbrace{(\boldsymbol{\rm h}_{{\rm d}, k}^H+\boldsymbol{\rm h}_{{\rm r}, k}^H\boldsymbol{\rm \Theta}_i\boldsymbol{\rm H})\boldsymbol{\rm v}_is_i}_{\textrm{The energy signal from the HAP}}\\
~~~~~+\underbrace{(h_{i,k}+\boldsymbol{\rm h}_{{\rm r},k}^H\boldsymbol{\rm \Theta}_i\boldsymbol{\rm g}_{{\rm r},i})\sqrt{p_{i}}x_i}_{\textrm{The uplink signal from other IoT devices}}+n_{k,i}^{\rm DL},
\end{array}
\end{equation}
where $\boldsymbol{\rm v}_i\in \mathbb{C}^{M_{\rm T}\times 1}$ denotes the energy beamforming vector in time slot $i$ with covariance matrix $\boldsymbol{\rm V}_i=\boldsymbol{\rm v}_i\boldsymbol{\rm v}_i^H$ and $\boldsymbol{\rm V}_i\succeq \boldsymbol{0}$. $s_i$ is a pseudo-random signal which is a prior known at the HAP satisfying $\mathbb{E}\{|s_i|^2\}=1$. $\boldsymbol{\rm h}_{{\rm d}, k}\in\mathbb{C}^{M_{\rm T}\times 1}$ and $\boldsymbol{\rm h}_{{\rm r}, k}\in\mathbb{C}^{N\times 1}$ denote the channel vectors from the HAP and the RIS to the $k$-th IoT device, respectively. $h_{i,k}$ denotes the channel coefficient from the $i$-th IoT device to IoT device $k$. $\boldsymbol{\rm g}_{{\rm r}, i}\in \mathbb{C}^{N\times 1}$ is the counterpart uplink channel vector. $\boldsymbol{\rm H}\in\mathbb{C}^{N\times M_{\rm T}}$ denotes the channel matrix from the HAP to the RIS. $\boldsymbol{\rm \Theta}_i \triangleq {\rm diag}(e^{j\theta_{i,1}},\cdot\cdot\cdot,e^{j\theta_{i,N}})$ is the diagonal phase-shift matrix in time slot $i$ during the WET phase, where $\theta_{i,n}\in [0,2\pi]$ being the corresponding phase shift. Based on the energy harvesting model, the harvested energy of the $k$-th IoT device is given by
\begin{equation}\label{eq36}
\begin{array}{l}
E_k^{\rm IoT}=\min\{\zeta E_{k}^{\rm IN},t_kP_{{\rm sat},k}^{\rm IoT}\},
\end{array}
\end{equation}
where $P_{{\rm sat},k}^{\rm IoT}$ denotes the saturation power of the $k$-th IoT device. $t_k$ denotes the $k$-th time slot. Furthermore, $E_{k}^{\rm IN}$ is
\begin{equation}\label{eq37}
\begin{array}{l}
E_{k}^{\rm IN}=\underbrace{\sum\limits_{i\not= k}^{K}t_i|(\boldsymbol{\rm h}_{{\rm d}, k}^H+\boldsymbol{\rm h}_{{\rm r}, k}^H\boldsymbol{\rm \Theta}_{i}\boldsymbol{\rm H})\boldsymbol{\rm v}_{i}|^2}_{\textrm{The received energy from the HAP}}\\
~~~~~+\underbrace{\sum\limits_{i\not= k}^{K}t_ip_{i}|h_{i,k}+\boldsymbol{\rm h}_{{\rm r},k}^H\boldsymbol{\rm \Theta}_i\boldsymbol{\rm g}_{{\rm r},i}|^2}_{\textrm{The received energy from other IoT devices}}.
\end{array}
\end{equation}
It is worth noting that we assume that IoT devices have certain initial energy so that the $1$-st IoT device can perform uplink WIT in the $1$-st time slot. Meanwhile, we adopt a periodic transmission protocol as shown in Fig. \ref{fig3} to provide enough energy to IoT devices and ensure stable transmission of IoT devices, i.e., the $k$-th IoT device can harvest energy during all time slots except for the $k$-th time slot, the harvested energy in the previous frame can also be used in the uplink WIT of the next frame\cite{g2}.
\subsection{The WIT Phase}
The $k$-th IoT device will transmit its own data to the HAP during the $k$-th time slot. Then the received signal at the HAP in the $k$-th time slot  is given by
\begin{equation}\label{eq38}
\begin{array}{l}
\boldsymbol{\rm y}_k^{\rm UL}=\underbrace{(\boldsymbol{\rm g}_{{\rm d}, k}+\boldsymbol{\rm G}^H\boldsymbol{\rm \Theta}_k\boldsymbol{\rm g}_{{\rm r}, k})\sqrt{p_k}x_k}_{\textrm{Desire signal}}\\
~~~~~+\underbrace{(\boldsymbol{\rm F}+\boldsymbol{\rm G}^H\boldsymbol{\rm \Theta}_k\boldsymbol{\rm H})\boldsymbol{\rm v}_ks_k}_{\textrm{Self-interference}}+\boldsymbol{\rm n}_k,
\end{array}
\end{equation}
where $x_k$ denotes the transmit signal of the $k$-th IoT device. $\boldsymbol{\rm g}_{{\rm d}, k}\in \mathbb{C}^{M_{\rm R}\times 1}$, $\boldsymbol{\rm G}\in \mathbb{C}^{N\times M_{\rm R}}$, and $\boldsymbol{\rm g}_{{\rm r}, k}\in \mathbb{C}^{N\times 1}$ are the counterpart uplink channels. $\boldsymbol{\rm F}\in \mathbb{C}^{M_{\rm R}\times M_{\rm T}}$ represents the effective loopback channel at
the HAP that satisfies $\mathbb{E}\{\|\boldsymbol{\rm F}\|_{\rm F}^2\}=\gamma$. $\boldsymbol{\rm n}_k\in\mathcal{CN}(\boldsymbol{0},\delta^2\boldsymbol{\rm I}_{M_{\rm R}})$ denotes the additive white Gaussian noise at the HAP. Similarly, we assume an imperfect SIC process, the quantization error after ADC can be modeled as an independent white Gaussian noise, i.e., $\boldsymbol{\rm n}_{{\rm error},k}\sim\mathcal{CN}(\boldsymbol{0}, \beta\sigma_{{\rm error},k}^2\boldsymbol{\rm I}_{M_{\rm R}})$, where $\beta\ll 1$ and $\sigma_{{\rm error},k}^2$ is given by
\begin{equation}\label{eq39}
\begin{array}{l}
\sigma_{{\rm error},k}^2\approx \gamma\|\boldsymbol{\rm v}_k\|^2.
\end{array}
\end{equation}

Considering that the transmission signals of IoT devices are usually very weak due to their battery and size limitations, a beamforming vector $\boldsymbol{\rm w}_k\in\mathbb{C}^{M_{\rm R}\times 1}$ is applied for enhancing the received signal from the $k$-th IoT device. Thus, the signal recovered at the HAP is given by
\begin{equation}\label{eq40}
\begin{array}{l}
\hat y_k{=}\boldsymbol{\rm w}_k^H(\boldsymbol{\rm g}_{{\rm d}, k}{+}\boldsymbol{\rm G}^H\boldsymbol{\rm \Theta}_k\boldsymbol{\rm g}_{{\rm r}, k})\sqrt{p_k}s_k{+}\boldsymbol{\rm w}_k^H\boldsymbol{\rm n}_{{\rm error},k}{+}\boldsymbol{\rm w}_k^H\boldsymbol{\rm n}_k.
\end{array}
\end{equation}

Then, the SINR of  the $k$-th IoT device’s signal recovered is given by
\begin{equation}\label{eq41}
\begin{array}{l}
\gamma_k=\frac{p_k|\boldsymbol{\rm w}_k^H(\boldsymbol{\rm g}_{{\rm d}, k}+\boldsymbol{\rm G}^H\boldsymbol{\rm \Theta}_k\boldsymbol{\rm g}_{{\rm r}, k})|^2}{\beta\gamma\|\boldsymbol{\rm v}_k\|^2+\delta^2}.
\end{array}
\end{equation}

Accordingly, the achievable throughput of the $k$-th IoT device is given by
\begin{equation}\label{eq42}
\begin{array}{l}
R_k=t_k\log_2(1+\gamma_k).
\end{array}
\end{equation}
\subsection{Problem Formulation}
In this subsection, we form an optimization problem that maximizes the sum throughput among all IoT devices. Mathematically, the problem is expressed as follows
\begin{equation}\label{eq43}
\begin{array}{l}
\max\limits_{\mbox{\scriptsize$\begin{array}{c} 
		\boldsymbol{\rm v}_i,\boldsymbol{\rm \Theta}_i,
		\boldsymbol{\rm w}_k,t_k,p_k,
		\end{array}$}} 
\sum\limits_{k=1}^KR_k\\
~~~~~~s.t.~{C_1}:~R_k\geq R_k^{\min},\\
~~~~~~~~~~~{C_2}:~t_kp_k+\sum\limits_{i\not=k}^Kp_{{\rm c},k}^{\rm passive}t_i+p_{{\rm c},k}^{\rm active}t_k\leq E_k^{\rm IoT},\\
~~~~~~~~~~~{C_3}:~\|\boldsymbol{\rm v}_k\|^2\leq P^{\max},\\
~~~~~~~~~~~{C_4}:~\left|[\boldsymbol{\rm \Theta}_k]_{n,n}\right|=1,\\
~~~~~~~~~~~{C_5}:~\sum\limits_{k=1}^Kt_k= T,t_k>0,\\
~~~~~~~~~~~{C_6}:~\|\boldsymbol{\rm w}_k\|^2=1,\\
\end{array}
\end{equation}
where $C_1$ denotes the minimum throughput constraint of each IoT device, $R_k^{\min}$ denotes the minimum throughput threshold of the $k$-th IoT device; $C_2$ denotes the total energy consumption of the $k$-th IoT device should not exceed the total harvested energy; $C_3$ denotes the maximum transmit power constraint of the HAP, $P^{\max}$ is the maximum transmit power threshold of the HAP; $C_4$ guarantees that the diagonal phase shift matrice $\boldsymbol{\rm \Theta}_k$ has $N$ unit modulus components on its main diagonal; $C_5$ denotes the total time constraint; $C_6$ denotes the receive beamforming constraint. 
\section{The Algorithm for User Switching-based Protocol}
It is noted that the proposed algorithm 1 cannot be applied to solve problem (\ref{eq43}) since there is no co-channel interference and the grouping factor in problem (\ref{eq43}), which motivates us to redesign a new algorithm to solve this problem in this subsection.

For any given $\boldsymbol{\rm v}_i$, $\boldsymbol{\rm \Theta}_i$, $t_k$, and $p_k$, it is well-known that the optimal receive beamforming design scheme is MRC, the receive beamforming is shown as
\begin{equation}\label{eq44}
\begin{array}{l}
\boldsymbol{\rm w}_k=\frac{\boldsymbol{\rm g}_{{\rm d}, k}+\boldsymbol{\rm G}^H\boldsymbol{\rm \Theta}_k\boldsymbol{\rm g}_{{\rm r}, k}}{\|\boldsymbol{\rm g}_{{\rm d}, k}+\boldsymbol{\rm G}^H\boldsymbol{\rm \Theta}_k\boldsymbol{\rm g}_{{\rm r}, k}\|}.
\end{array}
\end{equation}

With the MRC beamforming design, problem (\ref{eq43}) is equivalent to the following problem
\begin{equation}\label{eq45}
\begin{array}{l}
\max\limits_{\mbox{\scriptsize$\begin{array}{c} 
		\boldsymbol{\rm v}_i,\boldsymbol{\rm \Theta}_i,
		t_k,p_k,
		\end{array}$}} 
\sum\limits_{k=1}^Kt_k\log_2(1+\frac{p_k\|\boldsymbol{\rm g}_{{\rm d}, k}+\boldsymbol{\rm G}^H\boldsymbol{\rm \Theta}_k\boldsymbol{\rm g}_{{\rm r}, k}\|^2}{\beta\gamma\|\boldsymbol{\rm v}_k\|^2+\delta^2})\\
~~~~~~s.t.~{\bar C_1}:~t_k\log_2(1+\frac{p_k\|\boldsymbol{\rm g}_{{\rm d}, k}+\boldsymbol{\rm G}^H\boldsymbol{\rm \Theta}_k\boldsymbol{\rm g}_{{\rm r}, k}\|^2}{\beta\gamma\|\boldsymbol{\rm v}_k\|^2+\delta^2})\geq R_k^{\min},\\
~~~~~~~~~~~{C_2},{C_3},{C_4},{C_5}.
\end{array}
\end{equation}
It can be seen that problem (\ref{eq45}) is hard to solve due to the coupled relationship among optimization variables. In this regard, an AO method is proposed to solve this problem.
\subsubsection{Optimizing $p_k$ and $t_k$ for given $\boldsymbol{\rm v}_i$ and $\boldsymbol{\rm \Theta}_i$}
We define $\bar p_k=p_kt_k$, the optimization problem (\ref{eq45}) can be reformulated as the following problem
\begin{equation}\label{eq46}
\begin{array}{l}
\max\limits_{\mbox{\scriptsize$\begin{array}{c} 
		t_k,\bar p_k,
		\end{array}$}} 
\sum\limits_{k=1}^Kt_k\log_2(1+\frac{\bar p_k\|\boldsymbol{\rm g}_{{\rm d}, k}+\boldsymbol{\rm G}^H\boldsymbol{\rm \Theta}_k\boldsymbol{\rm g}_{{\rm r}, k}\|^2}{t_k(\beta\gamma\|\boldsymbol{\rm v}_k\|^2+\delta^2)})\\
~~s.t.~{C_5},{\tilde C_1}:~t_k\log_2(1{+}\frac{\bar p_k\|\boldsymbol{\rm g}_{{\rm d}, k}+\boldsymbol{\rm G}^H\boldsymbol{\rm \Theta}_k\boldsymbol{\rm g}_{{\rm r}, k}\|^2}{t_k(\beta\gamma\|\boldsymbol{\rm v}_k\|^2+\delta^2)}){\geq} R_k^{\min},\\
~~~~~~~{\bar C_2}:\bar p_k{+}\sum\limits_{i\not=k}^Kp_{{\rm c},k}^{\rm passive}t_i+p_{{\rm c},k}^{\rm active}t_k\leq \bar E_k^{\rm IoT},
\end{array}
\end{equation}
where
\begin{equation}\label{eq46-1}
\begin{array}{l}
\bar E_k^{\rm IoT}=\min\{\zeta (\sum\limits_{i\not= k}^{K}t_i|(\boldsymbol{\rm h}_{{\rm d}, k}^H+\boldsymbol{\rm h}_{{\rm r}, k}^H\boldsymbol{\rm \Theta}_{i}\boldsymbol{\rm H})\boldsymbol{\rm v}_{i}|^2\\
~~~~~~+\sum\limits_{i\not= k}^{K}\bar p_{i}|h_{i,k}+\boldsymbol{\rm h}_{{\rm r},k}^H\boldsymbol{\rm \Theta}_i\boldsymbol{\rm g}_{{\rm r},i}|^2),t_kP_{{\rm sat},k}^{\rm IoT}\}.
\end{array}
\end{equation}
Problem (\ref{eq46}) is a convex optimization problem and can be solved directly by the standard convex optimization
techniques.
\subsubsection{Optimizing $\boldsymbol{\rm v}_i$ and $\boldsymbol{\rm \Theta}_i$ for given $p_k$ and $t_k$} The optimization problem (\ref{eq45}) can be reduced as
\begin{equation}\label{eq47}
\begin{array}{l}
\max\limits_{\mbox{\scriptsize$\begin{array}{c} 
		\boldsymbol{\rm v}_i,\boldsymbol{\rm \Theta}_i,
		\end{array}$}} 
\sum\limits_{k=1}^Kt_k\log_2(1+\frac{p_k\|\boldsymbol{\rm g}_{{\rm d}, k}+\boldsymbol{\rm G}^H\boldsymbol{\rm \Theta}_k\boldsymbol{\rm g}_{{\rm r}, k}\|^2}{\beta\gamma\|\boldsymbol{\rm v}_k\|^2+\delta^2})\\
~~~~~~s.t.~{\bar C_1},{C_2}-{C_4}.
\end{array}
\end{equation}
Problem (\ref{eq47}) is a non-convex optimization problem due to the coupled relationship between $\boldsymbol{\rm v}_i$ and $\boldsymbol{\rm \Theta}_i$ in SINR and $E_k^{\rm IoT}$. Based on the SCA method, we substitute the numerators and the
denominators of $R_k$ by slack variables, i.e.,
\begin{equation}\label{eq48}
\begin{array}{l}
{C_{1-1}}:\beta\gamma\|\boldsymbol{\rm v}_k\|^2{+}\delta^2{+}p_k\|\boldsymbol{\rm g}_{{\rm d}, k}+\boldsymbol{\rm G}^H\boldsymbol{\rm \Theta}_k\boldsymbol{\rm g}_{{\rm r}, k}\|^2{\geq} e^{\alpha_k},
\end{array}
\end{equation}
\begin{equation}\label{eq49}
\begin{array}{l}
C_{1-2}:\beta\gamma\|\boldsymbol{\rm v}_k\|^2+\delta^2\leq e^{\beta_k}.
\end{array}
\end{equation}
To facilitate the design of an efficient algorithm, (\ref{eq48}) can be transformed into
\begin{equation}\label{eq50}
\begin{array}{l}
\bar C_{1-1}:~\beta\gamma\|\boldsymbol{\rm v}_k\|^2+\delta^2+p_k\|\boldsymbol{\rm g}_{{\rm d}, k}+\boldsymbol{\rm G}^H\boldsymbol{\rm \Theta}_k\boldsymbol{\rm g}_{{\rm r}, k}\|^2 \\
~~~~~\Rightarrow\beta\gamma\|\boldsymbol{\rm v}_k\|^2+\delta^2+p_k{\rm Tr}(\boldsymbol{{\rm G}}_k^H\boldsymbol{{\rm O}}_k\boldsymbol{{\rm G}}_k)\geq e^{\alpha_k},\\
\end{array}
\end{equation}
where $\boldsymbol{{\rm G}}_{k}=
[{\rm diag}(\boldsymbol{{\rm g}}_{{\rm r},k}^H)\boldsymbol{{\rm G}};
\boldsymbol{{\rm g}}_{{\rm d},k}^H]
$.

For the non-convexity of (\ref{eq49}), we apply the first order Taylor expansion, i.e.,
\begin{equation}\label{eq51}
\begin{array}{l}
\bar C_{1-2}:~\beta\gamma\|\boldsymbol{\rm v}_k\|^2+\delta^2\leq e^{\bar\beta_k}(\beta_k-\bar \beta_k+1).
\end{array}
\end{equation}

${C_{2}}$ is a non-convex constraint due to the coupled variables $\boldsymbol{\rm v}_i$ and $\boldsymbol{\rm \Theta}_i$. To this end, defining $\boldsymbol{{\rm V}}_k=\boldsymbol{{\rm v}}_k\boldsymbol{{\rm v}}_k^H$, which should satisfy constraints $\boldsymbol{{\rm V}}_k\succeq\boldsymbol{{0}}$, we have
\begin{equation}\label{eq52}
\begin{array}{l}
E_k^{\rm IoT}=\min\{\zeta(\sum\limits_{i\not= k}^{K}t_i|(\boldsymbol{\rm h}_{{\rm d}, k}^H+\boldsymbol{\rm h}_{{\rm r}, k}^H\boldsymbol{\rm \Theta}_{i}\boldsymbol{\rm H})\boldsymbol{\rm v}_{i}|^2\\
~~~~~~+\sum\limits_{i\not= k}^{K}t_ip_{i}|h_{i,k}+\boldsymbol{\rm h}_{{\rm r},k}^H\boldsymbol{\rm \Theta}_i\boldsymbol{\rm g}_{{\rm r},i}|^2),t_kP_{{\rm sat},k}^{\rm IoT}\}\\
~~~~~~= \min\{\zeta(\sum\limits_{i\not= k}^{K}t_i{\rm Tr}(\boldsymbol{\rm H}_k\boldsymbol{\rm V}_i\boldsymbol{\rm H}_k^H\boldsymbol{\rm O}_i)\\
~~~~~~+\sum\limits_{i\not= k}^{K}t_ip_{i}{\rm Tr}(\boldsymbol{\rm H}_{i,k}\boldsymbol{\rm O}_i)),t_kP_{{\rm sat},k}^{\rm IoT}\},\\
\end{array}
\end{equation}
where $\boldsymbol{{\rm o}}_i{=}
[\boldsymbol{{\rm \theta}}_{i};
1]$, $\boldsymbol{{\rm O}}_i{=}\boldsymbol{{\rm o}}_i\boldsymbol{{\rm o}}_i^H$, when ${\rm Rank}(\boldsymbol{{\rm O}}_i){=}1$ holds, $\boldsymbol{{\rm \Theta}}_i$ can be denoted by ${\rm diag}([\boldsymbol{{\rm O}}_i]_{N+1,1:N})$, where $[\boldsymbol{{\rm O}}_i]_{N+1,1:N}{=}\left[[\boldsymbol{{\rm O}}_i]_{N+1,1},\cdot\cdot\cdot,[\boldsymbol{{\rm O}}_i]_{N+1,N}\right]$. $\boldsymbol{{\rm H}}_{k}{=}[
{\rm diag}(\boldsymbol{{\rm h}}_{{\rm r},k}^H)\boldsymbol{{\rm H}};
\boldsymbol{{\rm h}}_{{\rm d},k}^H]$, $\boldsymbol{{\rm h}}_{i,k}{=}[
{\rm diag}(\boldsymbol{{\rm h}}_{{\rm r},k}^H)\boldsymbol{{\rm g}}_{{\rm r},i};
h_{i,k}]$, $\boldsymbol{{\rm H}}_{i,k}=\boldsymbol{{\rm h}}_{i,k}\boldsymbol{{\rm h}}_{i,k}^H$.

Based on the same method as (\ref{eq22}), we further rewrite the related terms as
\begin{equation}\label{eq53}
\begin{array}{l}
{\rm Tr}(\boldsymbol{\rm H}_k\boldsymbol{\rm V}_i\boldsymbol{\rm H}_k^H\boldsymbol{\rm O}_i)=\frac{1}{2}\|\boldsymbol{\rm O}_i+\boldsymbol{\rm H}_k\boldsymbol{\rm V}_i\boldsymbol{\rm H}_k^H\|_{\rm F}^2\\
~~~~~~~~~-\frac{1}{2}\|\boldsymbol{\rm O}_i\|_{\rm F}^2-\frac{1}{2}\|\boldsymbol{\rm H}_k\boldsymbol{\rm V}_i\boldsymbol{\rm H}_k^H\|_{\rm F}^2.
\end{array}
\end{equation}
Meanwhile, the term $\frac{1}{2}\|\boldsymbol{\rm O}_i+\boldsymbol{\rm H}_k\boldsymbol{\rm V}_i\boldsymbol{\rm H}_k^H\|_{\rm F}^2$ can be bounded by an affine function, i.e.,
\begin{equation}\label{eq54}
\begin{array}{l}
\frac{1}{2}\|\boldsymbol{\rm O}_i+\boldsymbol{\rm H}_k\boldsymbol{\rm V}_i\boldsymbol{\rm H}_k^H\|_{\rm F}^2\geq \frac{1}{2}\|\boldsymbol{\rm \bar O}_i+\boldsymbol{\rm H}_k\boldsymbol{\rm \bar V}_i\boldsymbol{\rm H}_k^H\|_{\rm F}^2\\
+{\rm Tr}\left((\boldsymbol{\rm \bar O}_i+\boldsymbol{\rm H}_k\boldsymbol{\rm \bar V}_i\boldsymbol{\rm H}_k^H)^H(\boldsymbol{\rm O}_i-\boldsymbol{\rm \bar O}_i)\right)\\
+{\rm Tr}\left((\boldsymbol{\rm H}_k^H\boldsymbol{\rm \bar O}_i\boldsymbol{\rm H}_k{+}\boldsymbol{\rm H}_k^H\boldsymbol{\rm H}_k\boldsymbol{\rm \bar V}_i\boldsymbol{\rm H}_k^H\boldsymbol{\rm H}_k)^H(\boldsymbol{\rm V}_i{-}\boldsymbol{\rm \bar V}_i)\right),
\end{array}
\end{equation}
where $\boldsymbol{\rm \bar V}_i$ and $\boldsymbol{\rm \bar O}_i$ are the previous iteration results of $\boldsymbol{\rm V}_i$ and $\boldsymbol{\rm O}_i$, respectively.
Thus, $C_{2}$ can be transformed into
\begin{equation}\label{eq55}
\begin{array}{l}
\tilde C_{2}:~t_k p_k{+}\sum\limits_{i\not=k}^Kp_{{\rm c},k}^{\rm passive}t_i+p_{{\rm c},k}^{\rm active}t_k\leq \tilde E_k^{\rm IoT},
\end{array}
\end{equation}
where
\begin{equation}\label{eq56}
\begin{array}{l}
\tilde E_k^{\rm IoT}=\zeta\min\{\sum\limits_{i\not= k}^{K}t_i(\frac{1}{2}\|\boldsymbol{\rm \bar O}_i+\boldsymbol{\rm H}_k\boldsymbol{\rm \bar V}_i\boldsymbol{\rm H}_k^H\|_{\rm F}^2\\
+{\rm Tr}\left((\boldsymbol{\rm \bar O}_i+\boldsymbol{\rm H}_k\boldsymbol{\rm \bar V}_i\boldsymbol{\rm H}_k^H)^H(\boldsymbol{\rm O}_i-\boldsymbol{\rm \bar O}_i)\right)\\
+{\rm Tr}\left((\boldsymbol{\rm H}_k^H\boldsymbol{\rm \bar O}_i\boldsymbol{\rm H}_k+\boldsymbol{\rm H}_k^H\boldsymbol{\rm H}_k\boldsymbol{\rm \bar V}_i\boldsymbol{\rm H}_k^H\boldsymbol{\rm H}_k)^H(\boldsymbol{\rm V}_i{-}\boldsymbol{\rm \bar V}_i)\right)\\
-\frac{1}{2}\|\boldsymbol{\rm O}_i\|_{\rm F}^2-\frac{1}{2}\|\boldsymbol{\rm H}_k\boldsymbol{\rm V}_i\boldsymbol{\rm H}_k^H\|_{\rm F}^2)\\
+\sum\limits_{i\not= k}^{K}t_ip_{i}{\rm Tr}(\boldsymbol{\rm H}_{i,k}\boldsymbol{\rm O}_i),t_kP_{{\rm sat},k}^{\rm IoT}\}.
\end{array}
\end{equation}

The penalty-based method as shown in (\ref{eq30}) can still be used to deal with the rank-one constraint. Then, the optimization problem can be transformed into
\begin{equation}\label{eq57}
\begin{array}{l}
\max\limits_{\mbox{\scriptsize$\begin{array}{c} 
		\boldsymbol{\rm V}_i,\boldsymbol{\rm O}_i,\alpha_k,\beta_k
		\end{array}$}} 
\sum\limits_{k=1}^K t_k\frac{1}{\ln2}(\alpha_k{-}\beta_k){-}\frac{1}{2\rho}(\|\boldsymbol{\rm O}_{k}\|_*{-}{ O}_{k})\\
s.t.~\bar C_{1-1}, \bar C_{1-2}, \tilde C_2, {\bar C_1}:~t_k\frac{1}{\ln2}(\alpha_k-\beta_k)\geq R_k^{\min},\\
~~~~~{\bar C_3}:~{\rm Tr}(\boldsymbol{\rm V}_k)\leq P^{\max},\boldsymbol{{\rm V}}_k\succeq\boldsymbol{{0}},\\
~~~~~{C_4}:~[\boldsymbol{{\rm O}}_k]_{n,n}\leq 1,\boldsymbol{{\rm O}}_k\succeq \boldsymbol{0},[\boldsymbol{{\rm O}}_k]_{N+1,N+1}=1,
\end{array}
\end{equation}
where ${ O}_{k}\triangleq \|\boldsymbol{\rm\bar O}_{k}\|_2+{\rm Tr}(\boldsymbol{{\rm\lambda}}_{2,\max}\boldsymbol{{\rm\lambda}}_{2,\max}^H(\boldsymbol{\rm O}_{k}-\boldsymbol{\rm\bar O}_{k}))$,  $\boldsymbol{{\rm\lambda}}_{2,\max}$ is the eigenvector associated with the principal eigenvalue of $\boldsymbol{\rm\bar O}_{k}$. $\rho>0$ is the penalty factor. It can be seen that problem (\ref{eq57}) is a convex optimization problem and can be solved directly by the standard convex optimization techniques. The user switching-based algorithm is shown in \textbf{Algorithm 2}.
\begin{spacing}{1.00}
	\floatname{algorithm}{Algorithm}
	\renewcommand{\algorithmicrequire}{\textbf{Input:}}
	\renewcommand{\algorithmicensure}{\textbf{Output:}}
	\begin{algorithm}[!t]
		\small
		\caption{The User Switching-based Algorithm }
		\begin{algorithmic}[1]
			\State Initialize system parameters: $K$, $N$, $T$, $P^{\max}$, $R_{k}^{\min}$, $\delta^2$, $P_{{\rm sat},k}^{\rm IoT}$, $p_{{\rm c},k}^{\rm active}$, $p_{{\rm c},k}^{\rm passive}$, $\beta$, $\gamma$, $\zeta$, $\boldsymbol{\rm\bar V}_{i}$,$\boldsymbol{\rm \bar O}_{i}$, $\alpha_{k}$, $\beta_{k}$;
			\State Set the maximum iteration number $L_{\max}$ and the convergence accuracy $\epsilon$, set the initial iteration index $ll=0$;
			\While{$ll\leq L_{\max}$}
			\State Initialize $\boldsymbol{{\rm \Theta}}_{i}$, $p_{k}$, $\boldsymbol{{\rm v}}_{i}$, and $t_{k}$;
			\State Calculate $\boldsymbol{{\rm w}}_{k}$ via (\ref{eq44});
			\State Given $\boldsymbol{{\rm w}}_{k}$, $\boldsymbol{{\rm \Theta}}_{k}$, and $\boldsymbol{{\rm v}}_{k}$, Calculate  $t_{k}$ and $p_{k}$ via (\ref{eq46});
			\Repeat
			\State Given $\boldsymbol{{\rm w}}_{k}$, $p_{k}$, and $t_{k}$, Calculate $\boldsymbol{{\rm O}}_{k}$ and $\boldsymbol{{\rm V}}_{k}$ via (\ref{eq57});
			\Until{converges};
			\State $\boldsymbol{{\rm \Theta}}_{k}={\rm diag}([\boldsymbol{{\rm O}}_{k}]_{N+1,1:N})$ and $\boldsymbol{{\rm v}}_{k}$ can be obtained by applying the eigenvalue decomposition method.
			\If{$|\sum\limits_{k=1}^KR_{k}(ll)-\sum\limits_{k=1}^KR_{k}(ll-1)|\leq\epsilon$}
			\State \textbf{break}.
			\Else
			\State $ll=ll+1$;
			\EndIf
			\EndWhile	
		\end{algorithmic}
	\end{algorithm}
\end{spacing}

\section{Performance Comparison}
In this section, to gain more insight into two proposed algorithms, we compare the uplink transmit power and the harvested energy of two protocols. In order to facilitate our comparison and analysis, we are restricted to the scenario where the whole frame length $T$ is evenly divided into $K$ time slots for the user switching-based protocol and $2K$ sub-time slots for the group switching-based protocol, and the phase shifts and channel environments for each IoT device under two algorithms are identical. Then, we have $h_{j,m}+\boldsymbol{\rm h}_{{\rm r},m}^H\boldsymbol{\rm \Theta}_{k,l}\boldsymbol{\rm g}_{{\rm r},j}=h_{i,k}+\boldsymbol{\rm h}_{{\rm r},k}^H\boldsymbol{\rm \Theta}_i\boldsymbol{\rm g}_{{\rm r},i}$ and $\boldsymbol{\rm h}_{{\rm d},m}^H+\boldsymbol{\rm h}_{{\rm r}, m}^H\boldsymbol{\rm \Theta}_{k,l}\boldsymbol{\rm H} =\boldsymbol{\rm h}_{{\rm d}, k}^H+\boldsymbol{\rm h}_{{\rm r}, k}^H\boldsymbol{\rm \Theta}_{i}\boldsymbol{\rm H}$. 

\subsection{The Transmit Power of IoT devices}
\textit{Proposition 1:} In the communication scenario above, the uplink transmit power under two protocols is identical when the number of IoT devices $K=2$. The uplink transmit power under the user switching-based protocol is greater than that under the group switching-based protocol when the number of IoT devices $K>2$, such as
\begin{equation}\label{eqA2}
\begin{array}{l}
\left\{
\begin{array}{l}
p_{m,k,\hat l}=p_k,~~ K=2,\\
p_{m,k,\hat l}<p_k,~~K>2.
\end{array}
\right.
\end{array}
\end{equation}
\begin{proof}
Please refer to Appendix A.
\end{proof}
\textit{Remark 1 :} Proposition 1 reveals the differences between the energy harvesting mechanisms under two protocols, the user switching-based protocol can harvest energy from the HAP during $\sum_{k=1}^{K-1}t_k$ time slots, while the group switching-based protocol can only harvest energy from the HAP during $\frac{\sum_{k=1}^{K}t_k}{2}$ time slots.
Thus, the time of energy harvested from the HAP under the user switching-based protocol is longer than that under the group switching-based protocol.

\subsection{The harvested energy of IoT devices}
\textit{Proposition 2:} In the communication scenario above, the energy harvested from the HAP and the energy recycled from other IoT devices under two protocols is identical when the number of IoT devices $K=2$. However, the energy harvested from the HAP and the energy recycled from other IoT devices under the user switching-based protocol is greater than that under the group switching-based protocol when the number of IoT devices $K>2$, such as
\begin{equation}\label{eqB2}
\begin{array}{l}
\left\{
\begin{array}{l}
E_{\rm GS}^{\rm EH}=E_{\rm US}^{\rm EH},~~ K=2,\\
E_{\rm GS}^{\rm EH}<E_{\rm US}^{\rm EH},~~K>2.
\end{array}
\right.
\end{array}
\end{equation}
where $E_{\rm GS}^{\rm EH}$ and $E_{\rm US}^{\rm EH}$ are the total harvested energy under the group switching-based protocol and the user switching-based protocol, respectively.
\begin{proof}
	Please refer to Appendix B.
\end{proof}
\textit{Remark 2:} Proposition 2 reveals that the energy harvested under the user switching-based protocol is greater than that under the group switching-based protocol. First, the time of energy harvested from the HAP under the user switching-based protocol is longer than that under the group switching-based protocol. Second, the user switching-based protocol is able to recycle energy from more IoT devices than the group switching-based protocol.

In general, for the communication scenarios of energy scarcity or the power sources with a low power level, we recommend the user switching-based protocol to provide more energy for IoT devices. For the communication scenarios with abundant energy or high throughput requirements, we recommend the group switching-based protocol to provide IoT devices with a more efficient WIT mechanism.

\section{Simulation Results}
\begin{figure}
	\centering
	\includegraphics[width=3.2in]{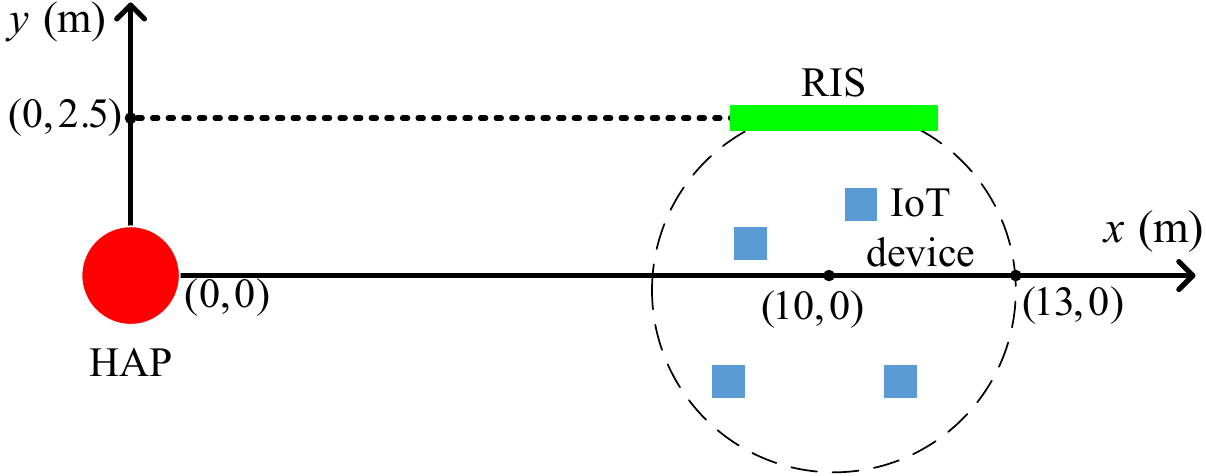}
	\caption{The simulated scenario of the considered RIS-aided WPCN.}
	\label{fig4}
\end{figure}
In this section, the effectiveness of the proposed algorithms is evaluated by comparing them with baseline algorithms. The group switching-based algorithm and the user switching-based algorithm are defined as the proposed algorithm 1 and the proposed algorithm 2, respectively, and baseline algorithms are defined as
\begin{itemize}
	\item \textbf{The proposed algorithms 1/2 without energy recycling}: The resource allocation algorithms for an RIS-aided WPCN are performed subject to the same constraint set as in (\ref{eq11}) or (\ref{eq43}), respectively, except that the energy recycling is not taken into account.
	\item \textbf{The proposed algorithms 1/2 without RIS}: The resource allocation algorithms for a conventional WPCN without RIS are performed subject to the same constraint set as in (\ref{eq11}) or (\ref{eq43}), respectively.
\end{itemize}
\begin{figure}
	\flushleft
	\vspace{-10pt}
	\begin{minipage}{0.8\linewidth}
		\centering
		\subfigure[The outer layer of the proposed algorithm 1]{
			\includegraphics[width=0.6\linewidth]{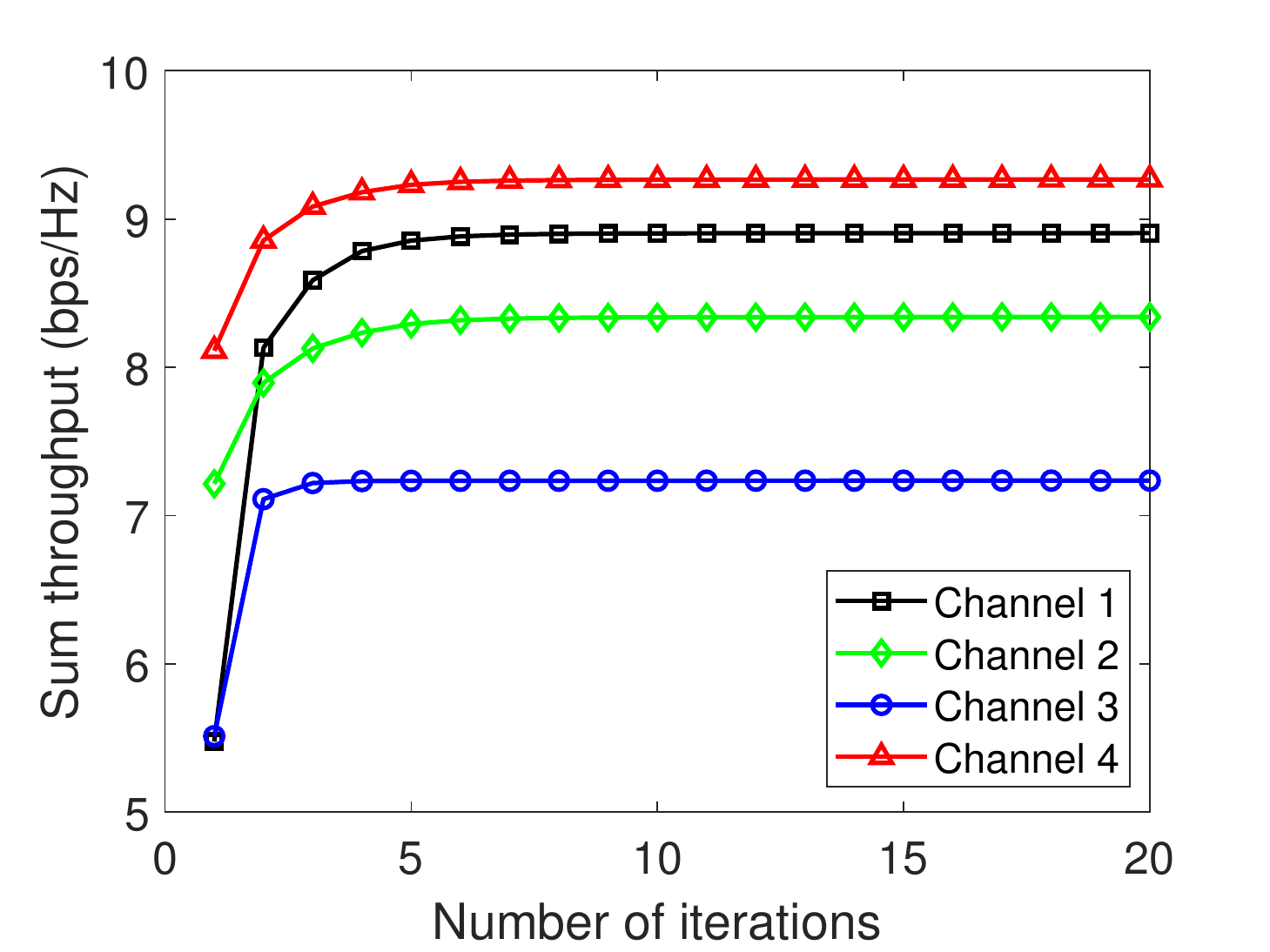}
			\label{fig5a}
		}\noindent
		\subfigure[The outer layer of the proposed algorithm 2]{
			\includegraphics[width=0.6\linewidth]{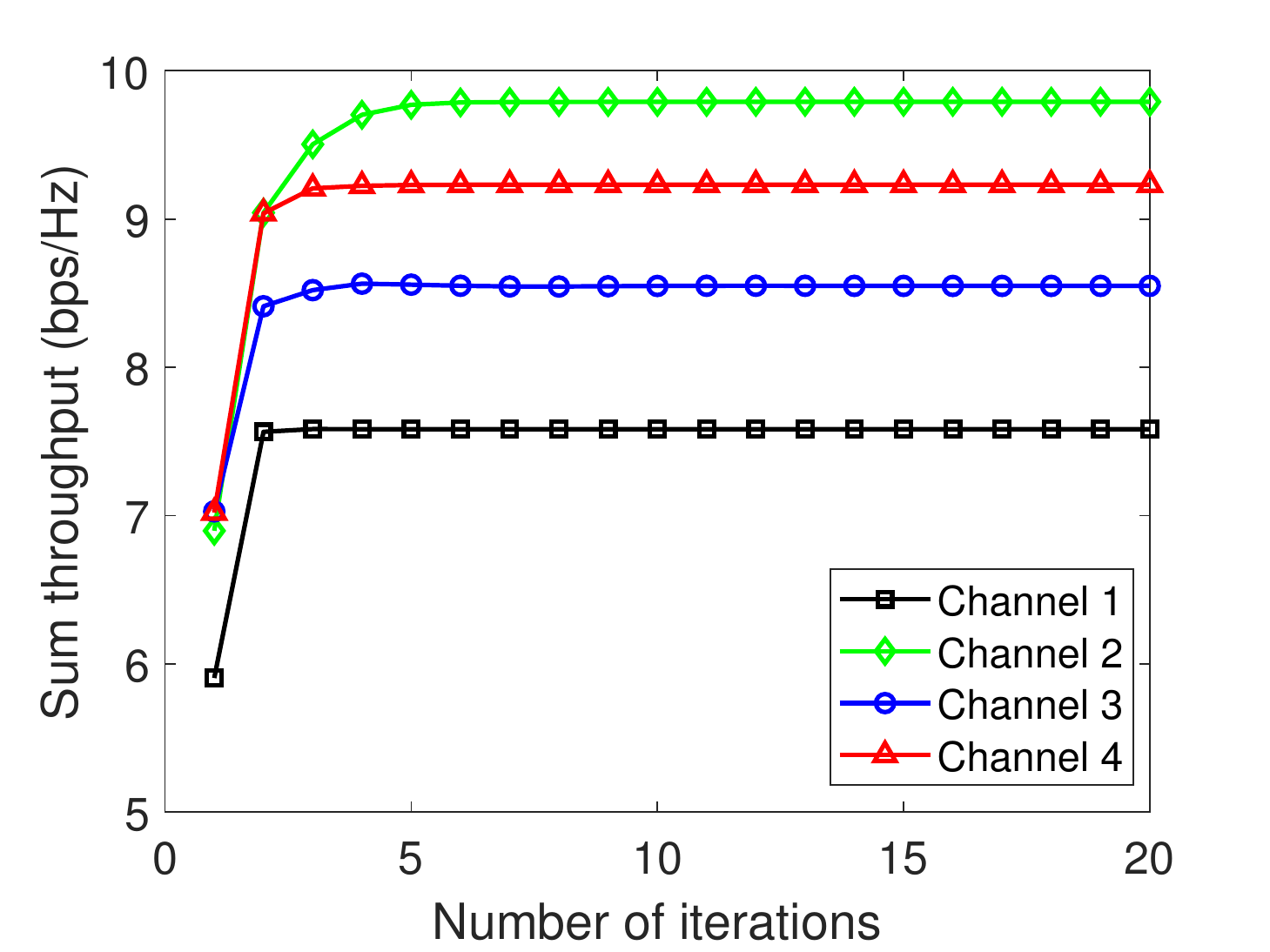}
			\label{fig5b}
		}
	\end{minipage}
	
	\begin{minipage}{0.8\linewidth}
		\centering
		\subfigure[The inner layer of the proposed algorithm 1]{
			\includegraphics[width=0.6\linewidth]{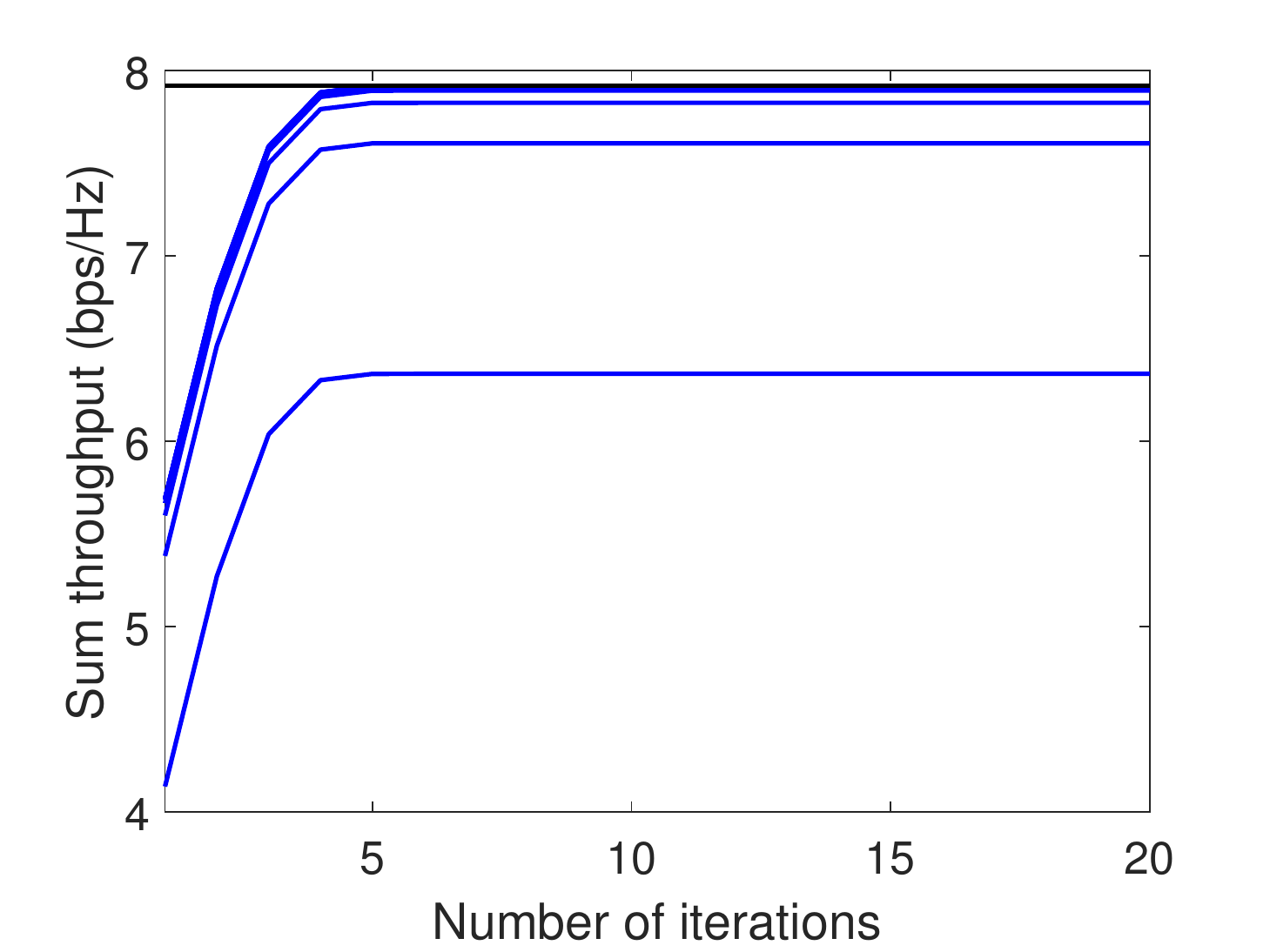}
			\label{fig5c}
		}\noindent
		\subfigure[The inner layer of the proposed algorithm 2]{
			\includegraphics[width=0.6\linewidth]{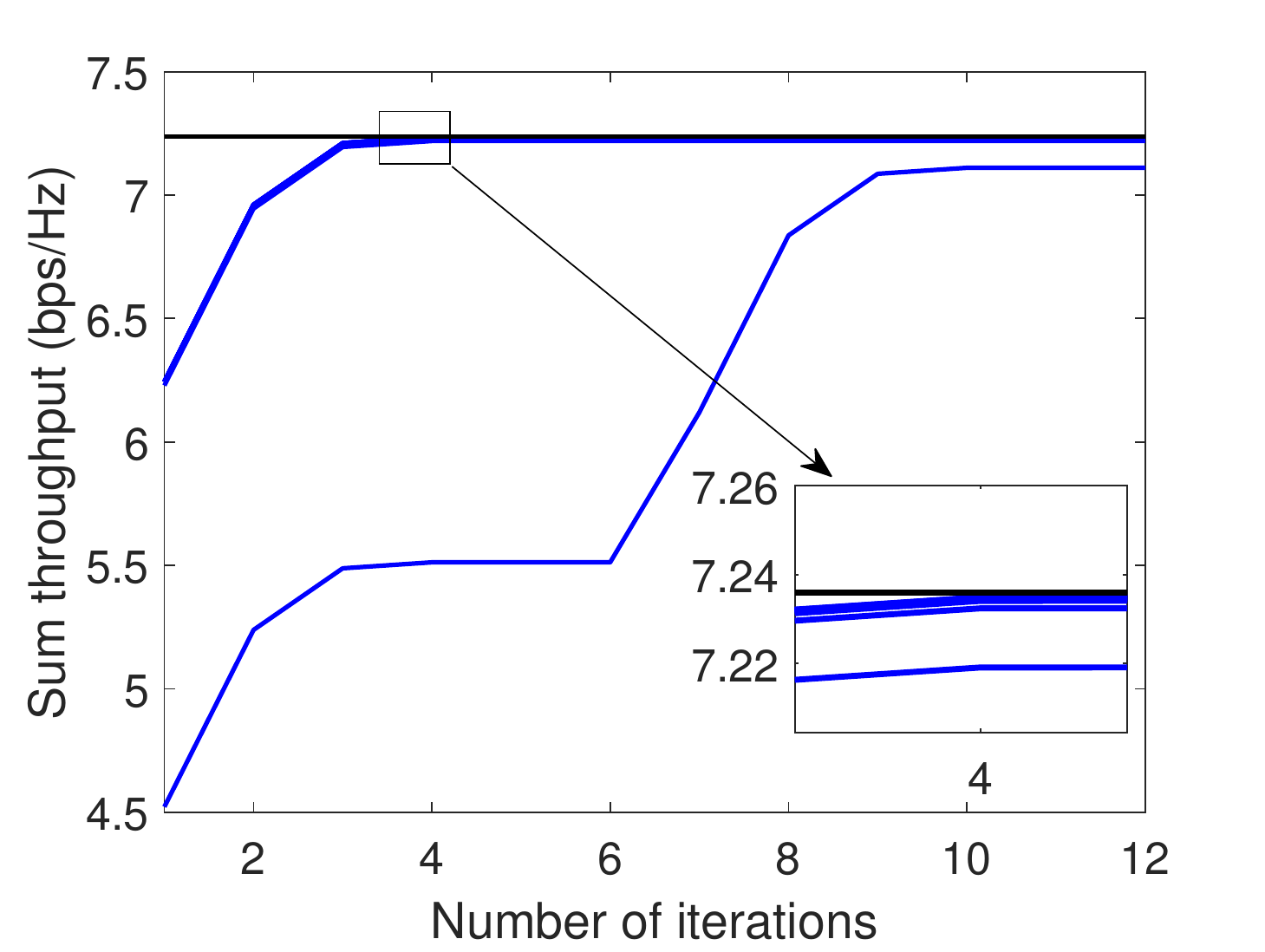}
			\label{fig5d}
		}
	\end{minipage}	
	\caption{Convergence analysis for proposed algorithms.}
\end{figure}
The schematic system model for the considered RIS-aided WPCN is shown in Fig. \ref{fig4}. A two-dimensional coordinate setup measured in meter (m) is considered, where the HAP and the RIS are located at $(0, 0)$ m, $(10, 2.5)$ m, while the IoT devices are uniformly and randomly distributed in a circle centered at $(10, 0)$ with a radius $3$ m. The path-loss mode is $A=A_0(\frac{d}{d_0})^{-\alpha}$, where $A_0=-30$ dB is the path-loss factor at $d_0=1$ m, $d$ is the distance between the transmitter and the receiver. The path-loss factors from the HAP-RIS links, the RIS-device links, the HAP-device links, and the device-device links are $2.2$, $2$, $2.6$, and $3$, respectively. The small-scale fading follows the Rayleigh distribution \cite{i4}. Other parameters are: $P^{\max}\in[30,40]$ dBm, $R_{k,m}^{\min}=R_{k}^{\min}=1$ bps/Hz, $\delta^2=-80$ dBm, $L^{\max}=10^4$, $P_{{\rm sat},k}^{\rm IoT}=29$ dBm. $p_{{\rm c},k}^{\rm active}=p_{{\rm c},m}^{\rm active}=p_{{\rm c},k}^{\rm passive}=p_{{\rm c},m}^{\rm passive}=-25$ dBm, $\beta=-60$ dB, $\gamma=-55$ dB, $\zeta=0.8$, $M_{\rm T}=M_{\rm R}=N=K=8$, $L_{\max}=10^3$, $\epsilon=10^{-3}$, and $T=1$.

\subsection{Performance Analysis of Proposed Algorithms}
Fig. 5 evaluates the convergence performance of the proposed algorithms 1 and 2. Fig. \ref{fig5a} and Fig. \ref{fig5b} shows the convergence of the outer layer iteration for the proposed algorithms 1 and 2 under four arbitrarily selected channel realizations. Fig. \ref{fig5c} and Fig. \ref{fig5d} show the convergence of the inner layer iteration for the proposed algorithms 1 and 2, where the black horizontal lines denote the values obtained by solving problem (\ref{eq11})/(\ref{eq43}) and the blue lines denote the convergence of each inner iteration. It is observed that each inner layer iteration converges to the  stable value with ten iterations, then the proposed algorithms 1 and 2 converge to the optimal points after several iterations of the inner layer.

\begin{figure}\vspace{-10pt}
	\centering
	\includegraphics[width=2.6in]{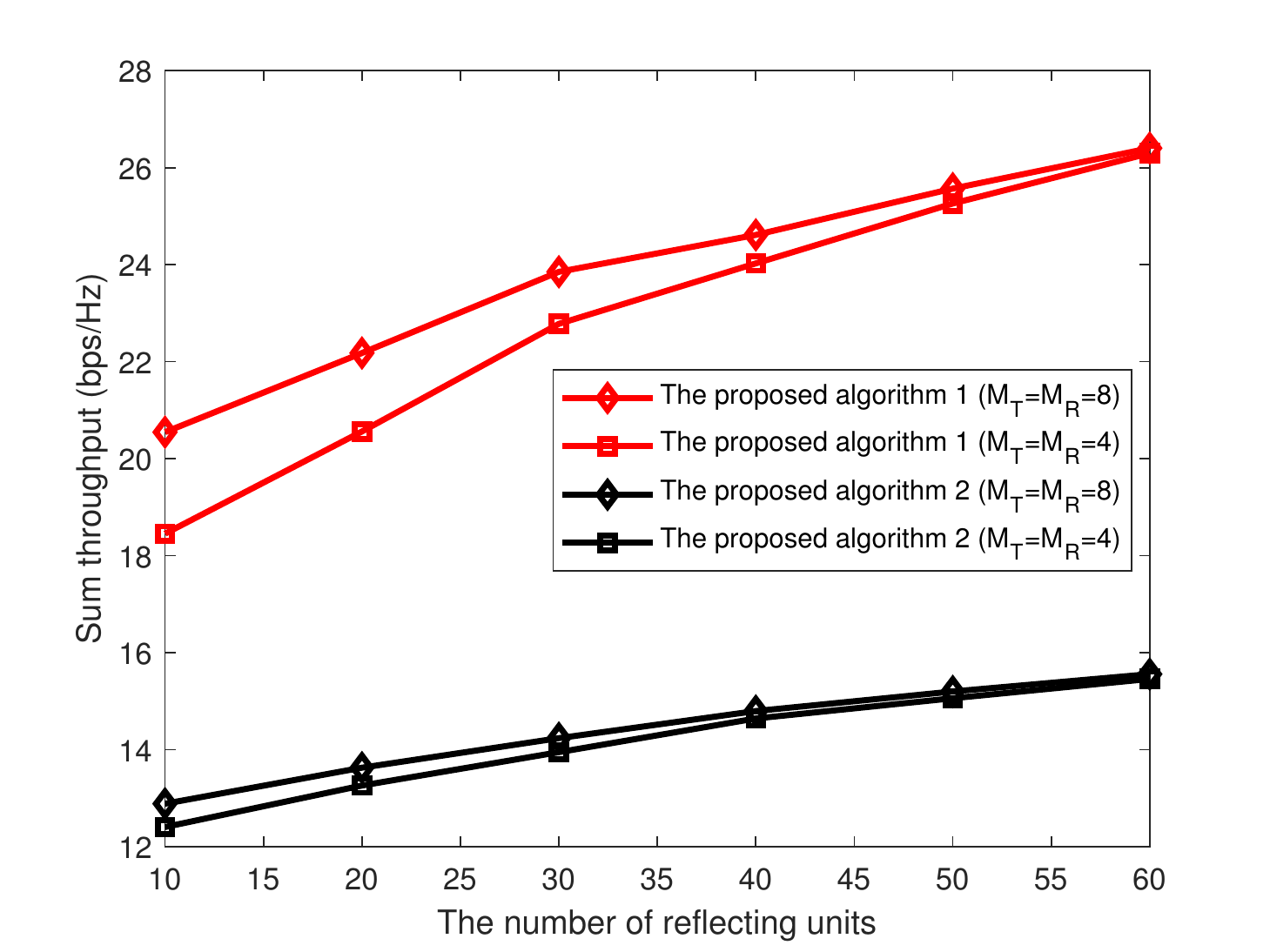}
	\caption{Sum throughput versus the number of reflecting units.}
	\label{fig6}
\end{figure}

Fig. \ref{fig6} illustrates the sum throughput of IoT devices versus the number of reflecting units $N$ under the proposed algorithms 1 and 2. From the figure, the sum throughput monotonically increases with the increasing $N$. This is because more reflecting units help achieve higher passive beamforming gain, which is beneficial for both uplink WIT and downlink WET. Besides, the sum throughput increases with the increasing number of antennas $M_T$ and $M_R$. In fact, the extra degrees of freedom offered by increasing the number of antennas can be exploited for more efficient resource allocation. Moreover, with $N$ increases, the performance gap between the different number of antennas gradually decreases for both the proposed algorithms 1 and 2 since the performance gain provided by the increased $N$ gradually compensates for the performance loss caused by the reduced $M_T$ and $M_R$.

\subsection{Performance Comparison of Proposed Algorithms}

\begin{figure}\vspace{-10pt}
	\centering
	\includegraphics[width=2.6in]{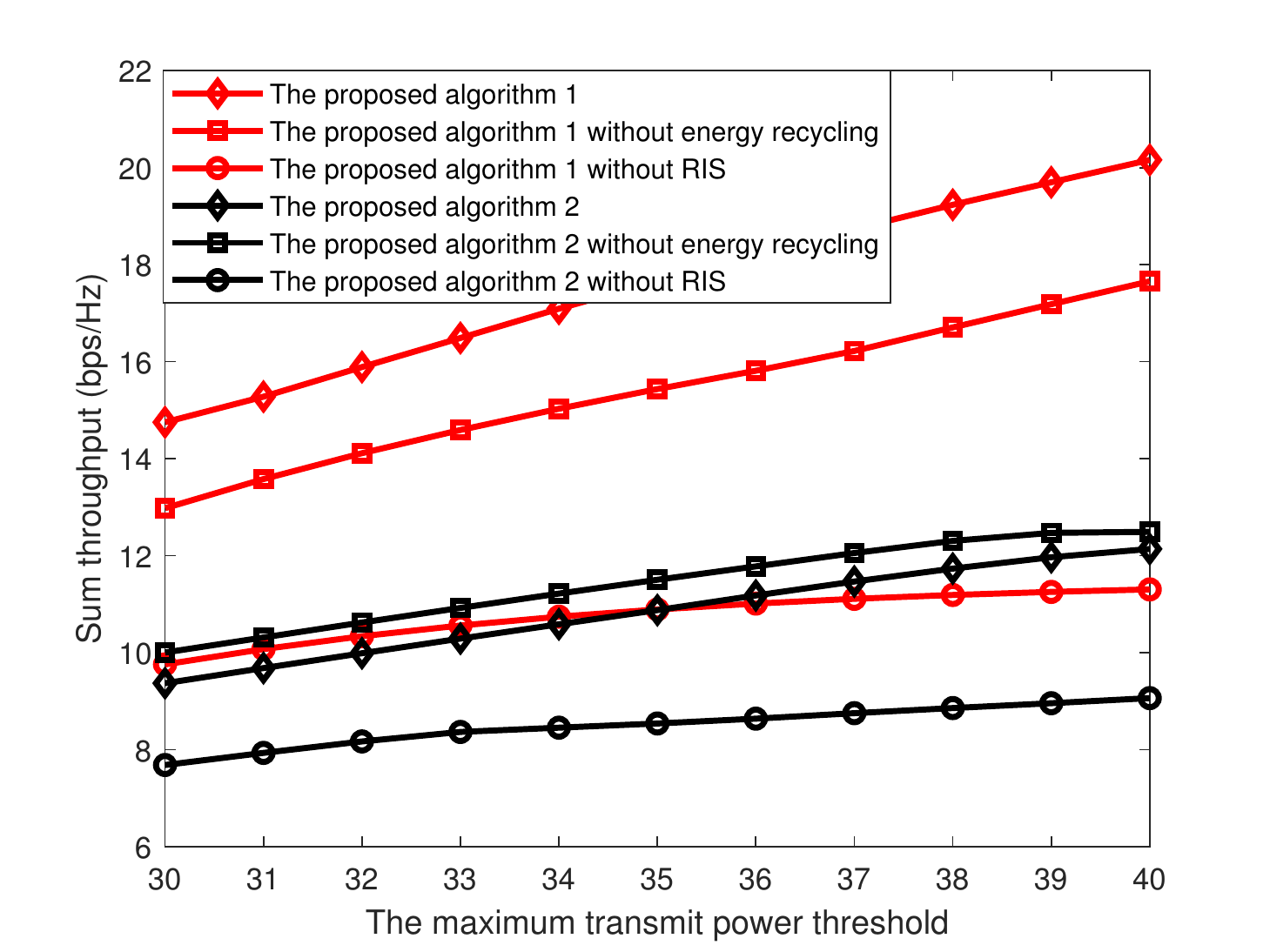}
	\caption{Sum throughput versus $P^{\max}$.}
	\label{fig7}
\end{figure}

Fig. \ref{fig7} illustrates the sum throughput of IoT devices versus the maximum transmit power threshold $P^{\max}$ under the proposed algorithms and the baseline algorithms. It can be observed that  the sum throughput under all algorithms increases when $P^{\max}$ increases. This is because an increase in $P^{\max}$ increases the energy harvested by IoT devices and the uplink transmit power of IoT devices, which effectively enlarges the sum throughput of IoT devices. In addition, the sum throughput under the proposed algorithm 1 is much higher than that of the proposed algorithm 2, which is caused by the different transmission mechanisms under two algorithms. Although the proposed algorithm 1 has co-channel interference in SINR, the efficiency of its transmission mechanism is much better than that of the proposed algorithm 1. Moreover, the sum throughput under corresponding baseline algorithms are  lower than the proposed algorithms 1 and 2. First, two baseline algorithms without energy recycling harvest less energy than the proposed algorithms 1 and 2 since they lack a part of the energy source, which results in lower uplink transmit power and lower in the sum throughput for two algorithms without energy recycling. Second, not surprisingly, the performance of the two baseline algorithms without RIS is the lowest due to the lack of reflective links, which also validates that the deployment of RIS can effectively improve the efficiency of uplink WIT.

\begin{figure}\vspace{-10pt}
	
	\centering
	\includegraphics[width=2.6in]{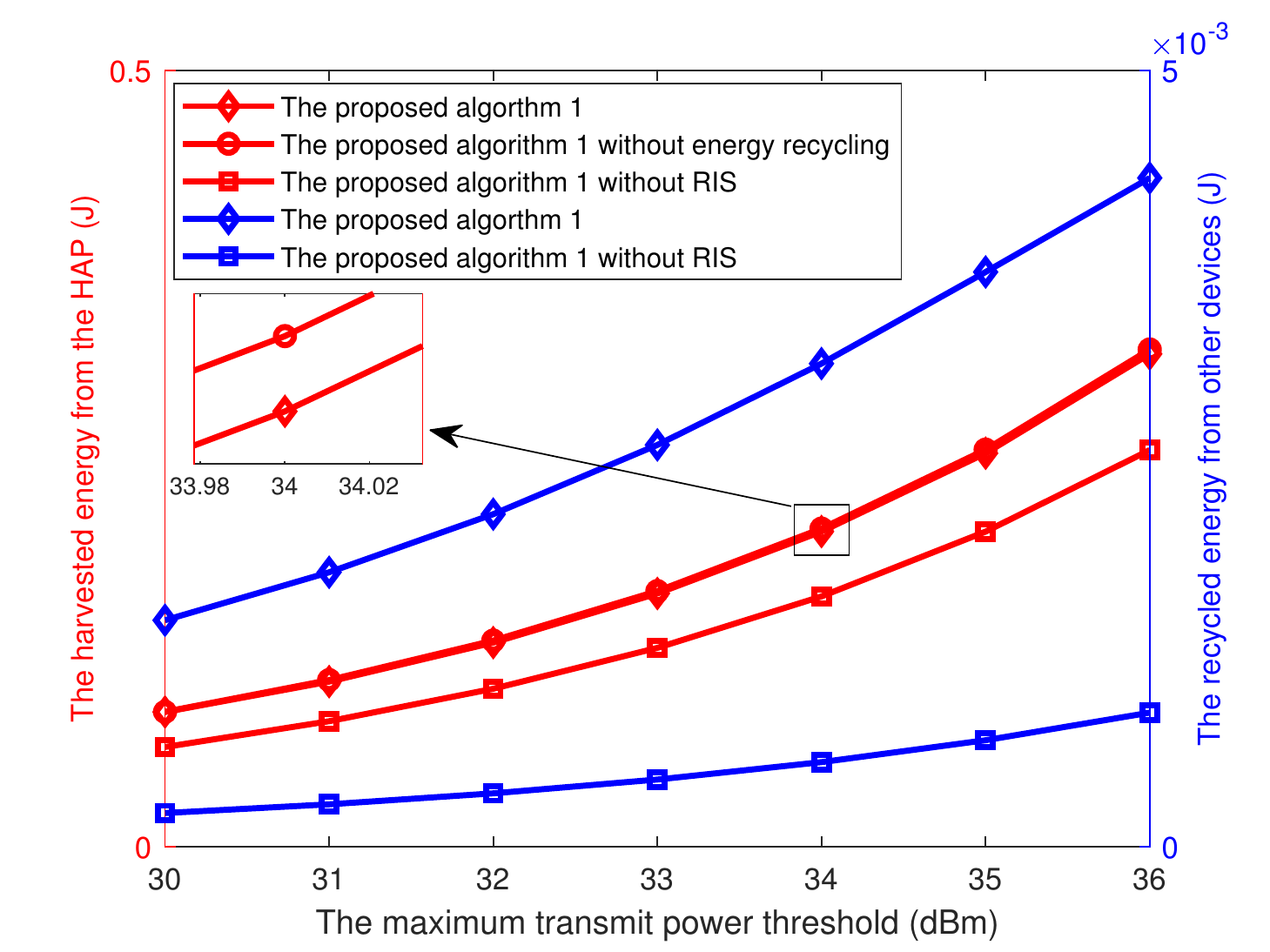}
	\caption{The harvested and recycled energy versus $P^{\max}$.}
	\label{fig8}
\end{figure}
Fig. \ref{fig8} illustrates the harvested energy from the HAP and the recycled energy from other devices versus $P^{\max}$ under the proposed algorithm 1 and the baseline algorithms. It can be clearly seen that the energy harvested from the HAP and the energy recycled from other IoT devices both increase with the increase of $P^{\max}$. According to (\ref{eq4}), the increase of $P^{\max}$ can increase the harvested energy, thus increasing the transmit power of the uplink WIT and also increasing the recycled energy. Interestingly, the energy harvested from the HAP under the proposed algorithm 1 without energy recycling is slightly higher than that of the proposed algorithm 1. The reason is that the proposed algorithm 1 without energy recycling lacks the recycled energy from other IoT devices. In order to increase the transmit power and throughput of uplink WIT, the system under the proposed algorithm 1 without energy recycling will adjust the phase shift of the RIS to favor the downlink WET, thus harvested more energy from the HAP than the proposed algorithm 1. However, the proposed algorithm 1 is inclined to adjust the phase shift to a compromise that favors both the uplink WIT and downlink WET due to the presence of the energy recycling-based mechanism. Moreover, less energy is harvested by the proposed algorithm 1 without RIS due to the large energy attenuation,  but this problem is alleviated by deploying the RIS. This result also verifies the effectiveness of the RIS.



\begin{figure}[!t]\vspace{-10pt}
	\centering
	\includegraphics[width=2.6in]{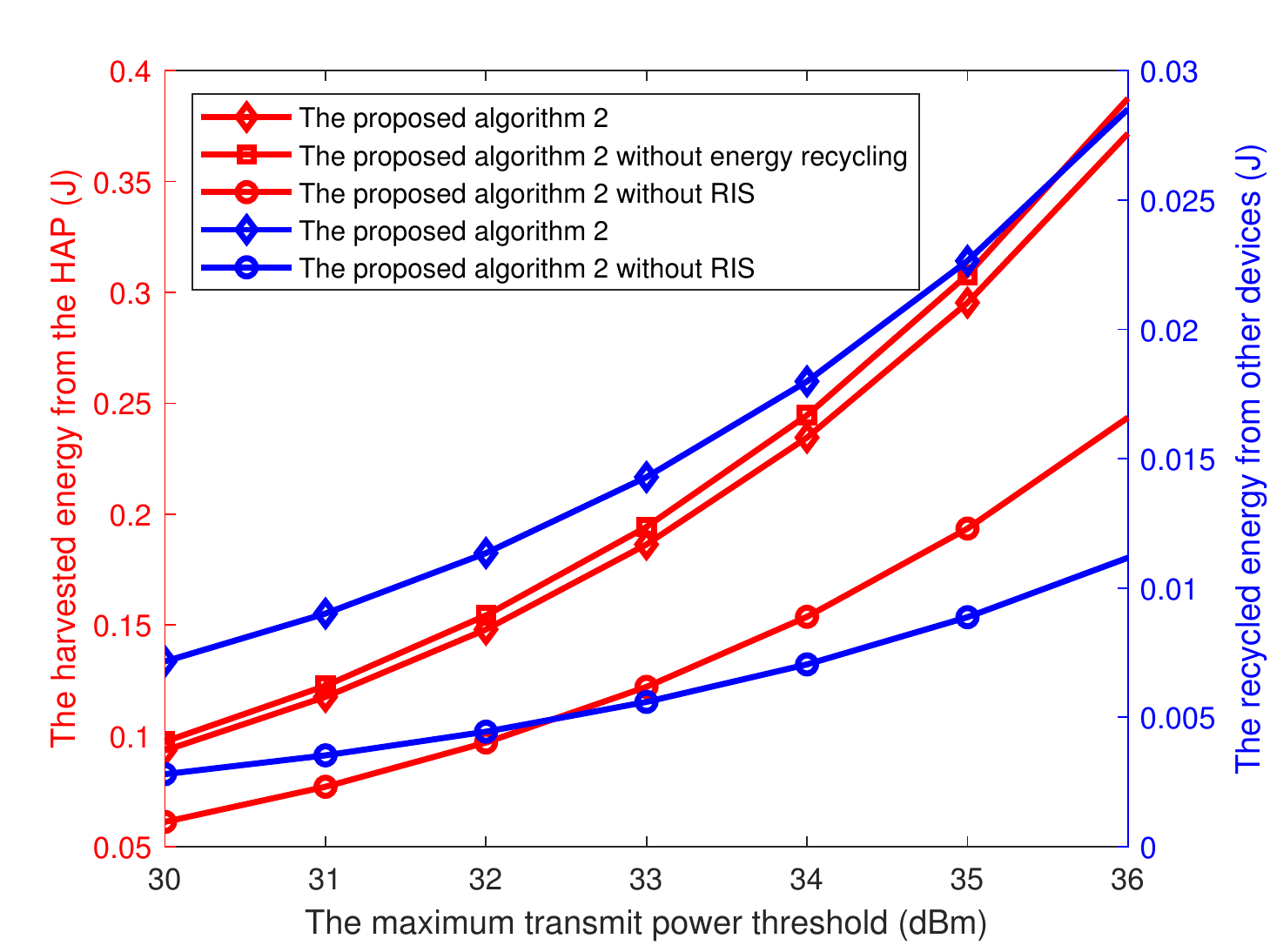}
	\caption{The harvested and recycled energy versus $P^{\max}$.}
	\label{fig9}
\end{figure}
Fig. \ref{fig9} illustrates the harvested energy from the HAP and the recycled energy from other devices versus $P^{\max}$ under the proposed algorithm 2 and the baseline algorithms. It can be observed that the energy harvested from the HAP and the energy recycled from other IoT devices both increase with the increase of $P^{\max}$. As illustrated in Fig. \ref{fig8}, according to (\ref{eq37}), the increase of $P^{\max}$ can increase the harvested energy and also increase the recycled energy. Besides, in order to increase the sum throughput, the proposed algorithm 2 without energy recycling adjusts the phase shift to try to harvest more energy from the HAP.

\begin{figure}
	\vspace{-10pt}
	\centering
	\includegraphics[width=2.6in]{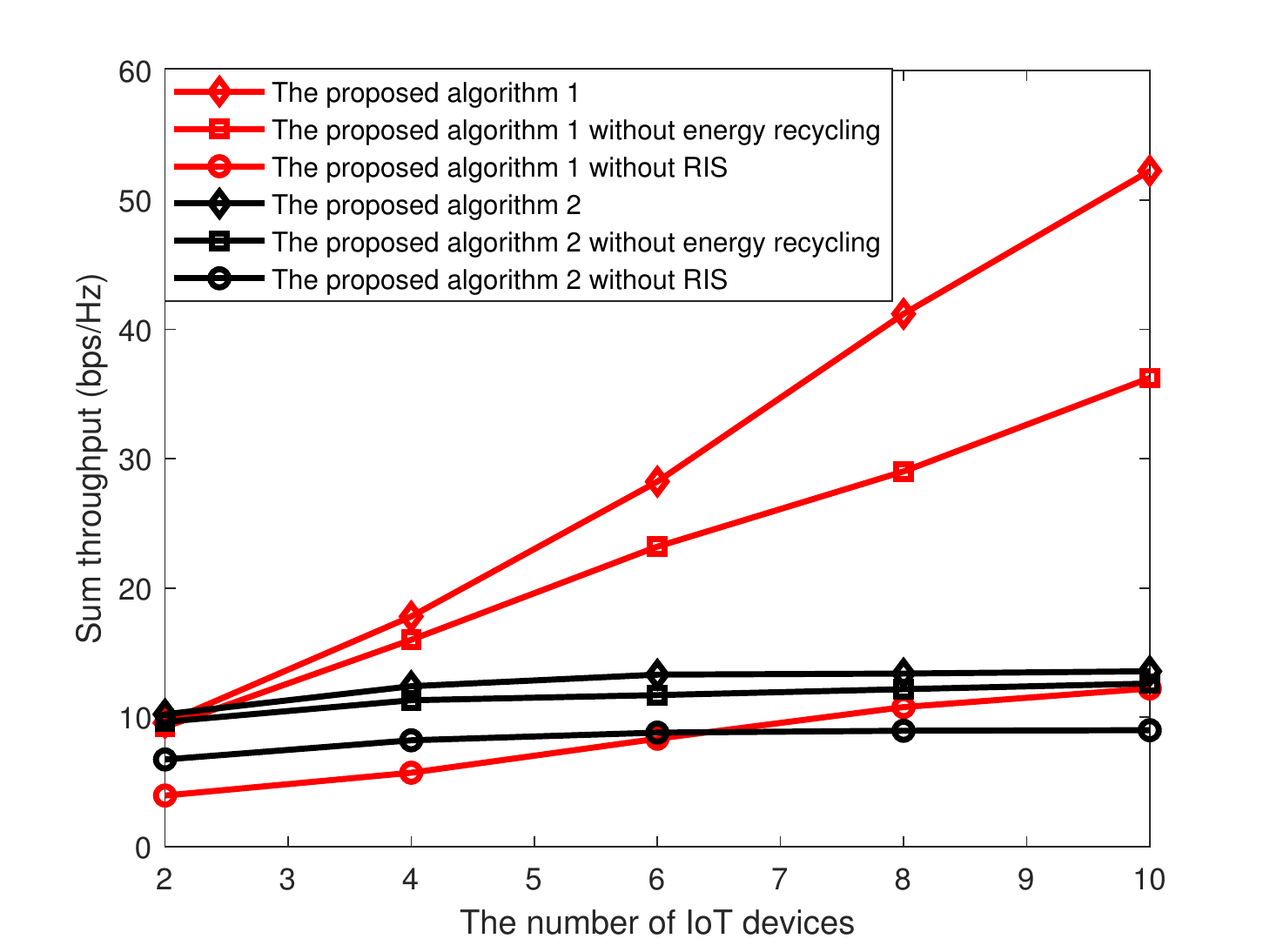}
	\caption{Sum throughput versus the number of IoT devices.}
	\label{fig10}
\end{figure}
Fig. \ref{fig10} illustrates the sum throughput versus the number of IoT devices under the proposed algorithms and baseline algorithms. It is obvious that the sum throughput under all algorithms increases as the increasing number of IoT devices since the system is guaranteed with minimum throughput constraints and more IoT devices participate in the uplink WIT process. In addition, several interesting insights are obtained. The performance gap between the proposed algorithms 1 and 2 are very small when the number of IoT devices is 2. This is because the transmission mechanisms of the two algorithms are similar when the number of IoT devices is 2. However, when the number of IoT devices increases, the sum throughput of the proposed algorithm 1 is much larger than that of the proposed algorithm 2. The reasons behind this are twofold: First, all IoT devices under the proposed algorithm 1 are involved in downlink WET and uplink WIT for each time slot, while only one IoT device is allowed to perform uplink WIT under the proposed algorithm 2. Second, as the number of IoT devices increases, the shortcomings of the proposed algorithm 2 become apparent. As the time slot allocated to each device becomes smaller and smaller, the throughput of each device may also decrease, which limits transmission performance. Due to the presence of the minimum throughput constraint, there is even the possibility of communication outage. The proposed algorithm 1 can overcome the above challenges well, and despite the presence of co-channel interference, the receive beamforming can better reduce the impact of co-channel interference on the system performance. These results verify the superiority of the group switching-based protocol.


\begin{figure}
	\centering
	\includegraphics[width=2.6in]{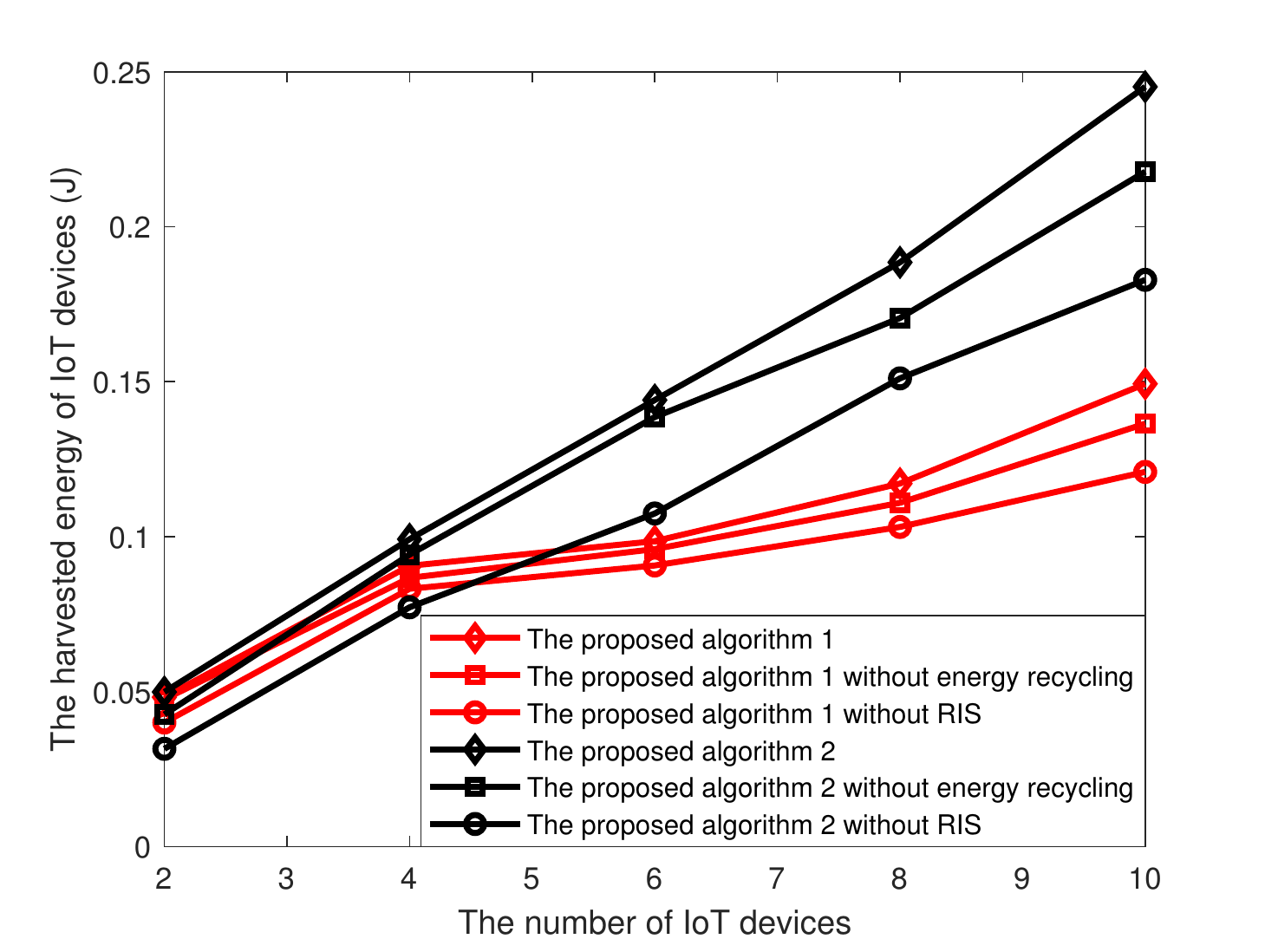}
	\caption{The total harvested energy versus the number of IoT devices.}
	\label{fig11}\vspace{-10pt}
\end{figure}
Fig. \ref{fig11} illustrates the total harvested energy of all IoT devices versus the number of IoT devices under the proposed algorithms and baseline algorithms. It can be observed that the total harvested energy of all IoT devices increases as the increasing number of IoT devices. In fact, a larger portion of the radiated power can be harvested when there are more IoT devices in the system since more IoT devices participate in the downlinks WET process. In addition, when the number of IoT devices is 2, the gap in the total harvested energy between the proposed algorithms 1 and 2 is very small, however, as the number of IoT devices increases, the energy harvested under the proposed algorithm 2 is higher than that of the proposed algorithm 1, which is caused by the difference in the WET mechanism, i.e, the proposed algorithm 2 can harvest energy from the HAP during $\sum_{k=1}^{K-1}t_k$ time slots, while the proposed algorithm 1 can only harvest energy from the HAP during $\frac{\sum_{k=1}^{K}t_k}{2}$ time slots. Moreover, the total harvested energy under the proposed algorithms is higher than that of the proposed algorithms without energy recycling. We relate to Fig. \ref{fig8} and \ref{fig9}, and find that although the proposed algorithms harvest less energy from the HAP, the total harvested energy is higher than that of the proposed algorithm without energy recycling. Therefore, the recycled energy from other devices is not negligible due to the intensive deployment of IoT devices. This result verifies the importance and effectiveness of energy recycling.

\section{Conclusions}
This paper investigates and analyzes energy recycling in an RIS-aided WPCN. Specially, the group switching- and user switching-based protocols are proposed to evaluate the impact of energy recycling on system performance. The corresponding problems are formulated with the objective of maximizing sum throughput while satisfying constraints on the maximum transmit power, the quality of service requirement, the phase-shift matrix, the transmission time, and the user grouping factor. Due to the intractable problems, we respectively develop two AO-based iterative algorithms applying the SCA method and the penalty-based method to deal with the highly non-convex problems. Simulation results demonstrate the effectiveness of energy recycling and the deployment of the RIS facilitates energy recycling. We also find that the group switching-based algorithm outperforms the user switching-based algorithm in terms of sum throughput, but the opposite behavior is observed in terms of harvested energy.

\appendices
\section{}
The sum throughput for these two protocols is monotonically increasing with respect to the uplink transmit power in the feasible range of transmit power when the self-interference is very small. The HAP will transmit using the maximum transmit power, meanwhile, IoT devices will exhaust all the harvested energy, thus increasing the uplink transmit power of IoT devices. Then, the downlink energy beamforming vectors of two protocols are identical since the channel environments of two protocols are identical, i.e., $\boldsymbol{\rm v}_{k,l}=\boldsymbol{\rm v}_{i}$. Then, we have
\begin{equation}\label{eq58}
\begin{array}{l}
t_{k,\hat l}p_{m,k,\hat l}+t_{k,l}p_{{\rm c},m}^{\rm passive}+t_{k,\hat l}p_{{\rm c},m}^{\rm active} = E_{m,k,l}^{\rm IoT}\\
\Rightarrow p_{m,k,\hat l}=\frac{\zeta |(\boldsymbol{\rm h}_{{\rm d},m}^H+\boldsymbol{\rm h}_{{\rm r}, m}^H\boldsymbol{\rm \Theta}_{k,l}\boldsymbol{\rm H})\boldsymbol{\rm v}_{k,l}|^2-p_{{\rm c},m}^{\rm passive}-p_{{\rm c},m}^{\rm active}}{1-\frac{K}{2}\zeta |h_{j,m}+\boldsymbol{\rm h}_{{\rm r},m}^H\boldsymbol{\rm \Theta}_{k,l}\boldsymbol{\rm g}_{{\rm r},j}|^2 },
\end{array}
\end{equation}
\begin{equation}\label{eq59}
\begin{array}{l}
t_kp_k+\sum\limits_{i\not=k}^Kp_{{\rm c},k}^{\rm passive}t_i+p_{{\rm c},k}^{\rm active}t_k = E_k^{\rm IoT}\\
\Rightarrow p_k = \frac{\zeta(K-1)|(\boldsymbol{\rm h}_{{\rm d}, k}^H+\boldsymbol{\rm h}_{{\rm r}, k}^H\boldsymbol{\rm \Theta}_{i}\boldsymbol{\rm H})\boldsymbol{\rm v}_{i}|^2-[(K-1)p_{{\rm c},k}^{\rm passive}+p_{{\rm c},k}^{\rm active}]}{1-\zeta(K-1)|h_{i,k}+\boldsymbol{\rm h}_{{\rm r},k}^H\boldsymbol{\rm \Theta}_i\boldsymbol{\rm g}_{{\rm r},i}|^2}.
\end{array}
\end{equation}

Furthermore, if we ignore the static power consumption of the circuit, then, we have the following inequality when $K\geq2$
\begin{equation}\label{eq60}
\small
\begin{array}{l}
p_{m,k,\hat l}=\frac{\zeta |(\boldsymbol{\rm h}_{{\rm d},m}^H+\boldsymbol{\rm h}_{{\rm r}, m}^H\boldsymbol{\rm \Theta}_{k,l}\boldsymbol{\rm H})\boldsymbol{\rm v}_{k,l}|^2}{1-\frac{K}{2}\zeta |h_{j,m}+\boldsymbol{\rm h}_{{\rm r},m}^H\boldsymbol{\rm \Theta}_{k,l}\boldsymbol{\rm g}_{{\rm r},j}|^2 }\leq \frac{\zeta |(\boldsymbol{\rm h}_{{\rm d},m}^H+\boldsymbol{\rm h}_{{\rm r}, m}^H\boldsymbol{\rm \Theta}_{k,l}\boldsymbol{\rm H})\boldsymbol{\rm v}_{k,l}|^2}{1-\zeta(K-1)|h_{i,k}+\boldsymbol{\rm h}_{{\rm r},k}^H\boldsymbol{\rm \Theta}_i\boldsymbol{\rm g}_{{\rm r},i}|^2 }\\
~~~~~~~\leq \frac{\zeta(K-1)|(\boldsymbol{\rm h}_{{\rm d}, k}^H+\boldsymbol{\rm h}_{{\rm r}, k}^H\boldsymbol{\rm \Theta}_{i}\boldsymbol{\rm H})\boldsymbol{\rm v}_{i}|^2}{1-\zeta(K-1)|h_{i,k}+\boldsymbol{\rm h}_{{\rm r},k}^H\boldsymbol{\rm \Theta}_i\boldsymbol{\rm g}_{{\rm r},i}|^2}=p_k.
\end{array}
\end{equation}
The proof is complete.

\section{}
For the group switching-based protocol, the harvested energy of all IoT devices during the whole frame length $T$ will be
\begin{equation}\label{eq61}
\begin{array}{l}
E_{{\rm GS}}^{\rm EH}=\zeta\sum\limits_{k= 1}^{K}\sum\limits_{m= 1}^{M}\sum\limits_{l= 1}^{2} t_{k,l}\{|(\boldsymbol{\rm h}_{{\rm d},m}^H+\boldsymbol{\rm h}_{{\rm r}, m}^H\boldsymbol{\rm \Theta}_{k,l}\boldsymbol{\rm H})\boldsymbol{\rm v}_{k,l}|^2\\
~~~~~+\sum\limits_{j\in G_{\hat{l}}}p_{j,k,l}|h_{j,m}+\boldsymbol{\rm h}_{{\rm r},m}^H\boldsymbol{\rm \Theta}_{k,l}\boldsymbol{\rm g}_{{\rm r},j}|^2\}\\
=\frac{K^2}{2}\zeta t_{k}|(\boldsymbol{\rm h}_{{\rm d},m}^H+\boldsymbol{\rm h}_{{\rm r}, m}^H\boldsymbol{\rm \Theta}_{k,l}\boldsymbol{\rm H})\boldsymbol{\rm v}_{k,l}|^2\\
~~~~~+\frac{K^3}{4}\zeta t_{k}p_{j,k,l}|h_{j,m}+\boldsymbol{\rm h}_{{\rm r},m}^H\boldsymbol{\rm \Theta}_{k,l}\boldsymbol{\rm g}_{{\rm r},j}|^2.
\end{array}
\end{equation}
Thus, we have
\begin{equation}\label{eq62}
\begin{array}{l}
E_{{\rm GS}}^{\rm EH}{=}
\left\{
\begin{array}{l}
\frac{K^2}{2}\zeta
t_{k}|(\boldsymbol{\rm h}_{{\rm d},m}^H{+}\boldsymbol{\rm h}_{{\rm r}, m}^H\boldsymbol{\rm \Theta}_{k,l}\boldsymbol{\rm H})\boldsymbol{\rm v}_{k,l}|^2,~~ K{=}1,\\
\frac{K^2}{2}\zeta t_{k}|(\boldsymbol{\rm h}_{{\rm d},m}^H+\boldsymbol{\rm h}_{{\rm r}, m}^H\boldsymbol{\rm \Theta}_{k,l}\boldsymbol{\rm H})\boldsymbol{\rm v}_{k,l}|^2\\
{+}\frac{K^3}{4}\zeta t_{k}p_{j,k,l}|h_{j,m}{+}\boldsymbol{\rm h}_{{\rm r},m}^H\boldsymbol{\rm \Theta}_{k,l}\boldsymbol{\rm g}_{{\rm r},j}|^2,~K{>}1.
\end{array}
\right.
\end{array}
\end{equation}

For the user switching-based protocol, the harvested energy of all IoT devices during the whole frame length $T$ will be
\begin{equation}\label{eq63}
\begin{array}{l}
E_{{\rm US}}^{\rm EH}=\sum\limits_{k= 1}^{K}\zeta\{\sum\limits_{i\not= k}^{K}t_i|(\boldsymbol{\rm h}_{{\rm d}, k}^H+\boldsymbol{\rm h}_{{\rm r}, k}^H\boldsymbol{\rm \Theta}_{i}\boldsymbol{\rm H})\boldsymbol{\rm v}_{i}|^2\\
~~~~~~+\sum\limits_{i\not= k}^{K}t_ip_{i}|h_{i,k}+\boldsymbol{\rm h}_{{\rm r},k}^H\boldsymbol{\rm \Theta}_i\boldsymbol{\rm g}_{{\rm r},i}|^2\}\\
=(K^2-K)\zeta t_k|(\boldsymbol{\rm h}_{{\rm d}, k}^H+\boldsymbol{\rm h}_{{\rm r}, k}^H\boldsymbol{\rm \Theta}_{i}\boldsymbol{\rm H})\boldsymbol{\rm v}_{i}|^2\\
~~~~~~+(K^2-K)\zeta t_kp_k|h_{i,k}+\boldsymbol{\rm h}_{{\rm r},k}^H\boldsymbol{\rm \Theta}_i\boldsymbol{\rm g}_{{\rm r},i}|^2.
\end{array}
\end{equation}

Futhermore, we ignore the static power consumption of the circuit since it is small and substitute (\ref{eq58}) and (\ref{eq59}) into (\ref{eq62}) and (\ref{eq63}), respectively.
\begin{equation}\label{eq64}
\begin{array}{l}
E_{{\rm GS}}^{\rm EH}=\frac{K^2}{2}\zeta t_{k}|(\boldsymbol{\rm h}_{{\rm d},m}^H+\boldsymbol{\rm h}_{{\rm r}, m}^H\boldsymbol{\rm \Theta}_{k,l}\boldsymbol{\rm H})\boldsymbol{\rm v}_{k,l}|^2\\
~~~~~~+\frac{K^3}{4}\zeta t_{k}p_{j,k,l}|h_{j,m}+\boldsymbol{\rm h}_{{\rm r},m}^H\boldsymbol{\rm \Theta}_{k,l}\boldsymbol{\rm g}_{{\rm r},j}|^2\\
\approx\frac{K^2}{2}\zeta t_{k}|(\boldsymbol{\rm h}_{{\rm d},m}^H+\boldsymbol{\rm h}_{{\rm r}, m}^H\boldsymbol{\rm \Theta}_{k,l}\boldsymbol{\rm H})\boldsymbol{\rm v}_{k,l}|^2\\
+\frac{\frac{K^3}{4}\zeta^2 t_{k}|h_{j,m}+\boldsymbol{\rm h}_{{\rm r},m}^H\boldsymbol{\rm \Theta}_{k,l}\boldsymbol{\rm g}_{{\rm r},j}|^2 |(\boldsymbol{\rm h}_{{\rm d},m}^H+\boldsymbol{\rm h}_{{\rm r}, m}^H\boldsymbol{\rm \Theta}_{k,l}\boldsymbol{\rm H})\boldsymbol{\rm v}_{k,l}|^2}{1-\frac{K}{2}\zeta |h_{j,m}+\boldsymbol{\rm h}_{{\rm r},m}^H\boldsymbol{\rm \Theta}_{k,l}\boldsymbol{\rm g}_{{\rm r},j}|^2 },\\
\end{array}
\end{equation}
\begin{equation}\label{eq65}
\begin{array}{l}
E_{{\rm US}}^{\rm EH}=(K^2-K)\zeta t_k|(\boldsymbol{\rm h}_{{\rm d}, k}^H+\boldsymbol{\rm h}_{{\rm r}, k}^H\boldsymbol{\rm \Theta}_{i}\boldsymbol{\rm H})\boldsymbol{\rm v}_{i}|^2\\
~~~~~~+(K^2-K)\zeta t_kp_k|h_{i,k}+\boldsymbol{\rm h}_{{\rm r},k}^H\boldsymbol{\rm \Theta}_i\boldsymbol{\rm g}_{{\rm r},i}|^2\\
\approx(K^2-K)\zeta t_k|(\boldsymbol{\rm h}_{{\rm d}, k}^H+\boldsymbol{\rm h}_{{\rm r}, k}^H\boldsymbol{\rm \Theta}_{i}\boldsymbol{\rm H})\boldsymbol{\rm v}_{i}|^2\\
+\frac{K(K-1)^2\zeta^2 t_k|h_{i,k}+\boldsymbol{\rm h}_{{\rm r},k}^H\boldsymbol{\rm \Theta}_i\boldsymbol{\rm g}_{{\rm r},i}|^2|(\boldsymbol{\rm h}_{{\rm d}, k}^H+\boldsymbol{\rm h}_{{\rm r}, k}^H\boldsymbol{\rm \Theta}_{i}\boldsymbol{\rm H})\boldsymbol{\rm v}_{i}|^2}{1-\zeta(K-1)|h_{i,k}+\boldsymbol{\rm h}_{{\rm r},k}^H\boldsymbol{\rm \Theta}_i\boldsymbol{\rm g}_{{\rm r},i}|^2}.\\
\end{array}
\end{equation}
Then, when $K\geq2$, we have
\begin{equation}\label{eq66-0}
\begin{array}{l}
\frac{K^2}{2}\zeta t_{k}|(\boldsymbol{\rm h}_{{\rm d},m}^H+\boldsymbol{\rm h}_{{\rm r}, m}^H\boldsymbol{\rm \Theta}_{k,l}\boldsymbol{\rm H})\boldsymbol{\rm v}_{k,l}|^2\\
\leq (K^2-K)\zeta t_k|(\boldsymbol{\rm h}_{{\rm d}, k}^H+\boldsymbol{\rm h}_{{\rm r}, k}^H\boldsymbol{\rm \Theta}_{i}\boldsymbol{\rm H})\boldsymbol{\rm v}_{i}|^2,
\end{array}
\end{equation}

\begin{equation}\label{eq66}
\begin{array}{l}
\frac{\frac{K^3}{4}\zeta^2 t_{k}|h_{j,m}+\boldsymbol{\rm h}_{{\rm r},m}^H\boldsymbol{\rm \Theta}_{k,l}\boldsymbol{\rm g}_{{\rm r},j}|^2 |(\boldsymbol{\rm h}_{{\rm d},m}^H+\boldsymbol{\rm h}_{{\rm r}, m}^H\boldsymbol{\rm \Theta}_{k,l}\boldsymbol{\rm H})\boldsymbol{\rm v}_{k,l}|^2}{1-\frac{K}{2}\zeta |h_{j,m}+\boldsymbol{\rm h}_{{\rm r},m}^H\boldsymbol{\rm \Theta}_{k,l}\boldsymbol{\rm g}_{{\rm r},j}|^2 }\\
\leq \frac{\frac{K^3}{4}\zeta^2 t_{k}|h_{j,m}+\boldsymbol{\rm h}_{{\rm r},m}^H\boldsymbol{\rm \Theta}_{k,l}\boldsymbol{\rm g}_{{\rm r},j}|^2 |(\boldsymbol{\rm h}_{{\rm d},m}^H+\boldsymbol{\rm h}_{{\rm r}, m}^H\boldsymbol{\rm \Theta}_{k,l}\boldsymbol{\rm H})\boldsymbol{\rm v}_{k,l}|^2}{1-(K-1)\zeta |h_{j,m}+\boldsymbol{\rm h}_{{\rm r},m}^H\boldsymbol{\rm \Theta}_{k,l}\boldsymbol{\rm g}_{{\rm r},j}|^2 }\\
\leq \frac{K(K-1)^2\zeta^2 t_k|h_{i,k}+\boldsymbol{\rm h}_{{\rm r},k}^H\boldsymbol{\rm \Theta}_i\boldsymbol{\rm g}_{{\rm r},i}|^2|(\boldsymbol{\rm h}_{{\rm d}, k}^H+\boldsymbol{\rm h}_{{\rm r}, k}^H\boldsymbol{\rm \Theta}_{i}\boldsymbol{\rm H})\boldsymbol{\rm v}_{i}|^2}{1-\zeta(K-1)|h_{i,k}+\boldsymbol{\rm h}_{{\rm r},k}^H\boldsymbol{\rm \Theta}_i\boldsymbol{\rm g}_{{\rm r},i}|^2}.
\end{array}
\end{equation}
The proof is complete.

\end{document}